\documentclass[12pt]{article}
\pdfoutput=1
\usepackage{amsmath,amssymb,amsfonts}
\usepackage{graphicx}
\usepackage{booktabs,colortbl}
\usepackage{cancel,xcolor}
\usepackage[normal]{caption}
\usepackage{subcaption}
\usepackage{authblk}

\usepackage{ulem}

\usepackage{color}

\newcommand{\be}{\begin{equation}} 
\newcommand{\ee}{\end{equation}} 
\newcommand{\bea}{\begin{eqnarray}}  
\newcommand{\eea}{\end{eqnarray}}
\newcommand{\bs}{\begin{split}} 
\newcommand{\es}{\end{split}}

\newcommand{\units}[1]{~\mathrm{#1}}

\newcommand{\ctoprule}{\toprule[0.5mm]}
\newcommand{\cbottomrule}{\bottomrule[0.5mm]}

\newcommand{\cmrule}{\midrule[0.25mm]}

\newcommand{\tr}{\operatorname{tr}}
\newcommand{\TSspc}{\!\!\!\!\!\!\!\!\!\!\!}

\begin{document}

\begin{flushright}
CERN-PH-TH/2013-263\\
UAB-FT-747\\
\end{flushright}
\vspace*{5mm}

\renewcommand{\thefootnote}{\fnsymbol{footnote}}
\setcounter{footnote}{1}
\begin{center}
{\Large \sc Indirect Effects of Supersymmetric Triplets in Stop Decays}\\
\vspace*{0.75cm}
{\textbf{J.\ de Blas}~\footnote{E-mail: Jorge.DeBlasMateo@roma1.infn.it}$^{1,2}$},
{\textbf{A.\ Delgado}~\footnote{E-mail: antonio.delgado@nd.edu}$^{2}$},
{\textbf{B.\ Ostdiek}~\footnote{E-mail: bostdiek@nd.edu}$^{2}$},
and
{\textbf{M.\ Quir\'os}~\footnote{E-mail: quiros@ifae.es}$^{3,4}$}\\
\vspace{0.5cm}
${}^1${\it INFN, Sezione di Roma, Piazzale A. Moro 2, I-00185 Rome, Italy}\\
\vspace{.2cm}
${}^2${\it Department of Physics, University of Notre Dame, Notre Dame, IN 46556, USA}\\
\vspace{.2cm}
${}^3${\it Dep. of Physics, CERN-TH Division, CH-1211, Geneva 23, Switzerland}\\
\vspace{.2cm}
${}^4${\it Instituci\'o Catalana de Recerca i Estudis  
Avan\c{c}ats (ICREA) and\\ Institut de F\'isica d'Altes Energies, Universitat Aut{\`o}noma de Barcelona\\
08193 Bellaterra, Barcelona, Spain}\\
\end{center}
\vspace{.5cm}

\begin{abstract}
 
\noindent We study an extension of the minimal supersymmetric standard model with a zero hypercharge triplet, and the effect that such a particle has on stop decays.
This model has the capability of predicting a 125.5 GeV Higgs even in the presence of light stops and it can modify the diphoton rate by means of the extra charged fermion triplet coupled to the Higgs. Working in the limit where the scalar triplet decouples, and with small values of $m_A$, we find that the fermion triplet can greatly affect the branching ratios of the stops, even in the absence of a direct stop-triplet coupling. We compare the triplet extension with the MSSM and discuss how the additional fields affect the search for stop pair production.

\end{abstract}.

\renewcommand{\thefootnote}{\arabic{footnote}}
\setcounter{footnote}{0}

\newpage
\section{Introduction}
\label{section_Intro}

The  ATLAS and CMS collaborations at CERN are trying to determine whether the discovered boson is the expected Standard Model (SM) Higgs~\cite{Aad:2012tfa, Chatrchyan:2012ufa}. They do this by examining its couplings to other particles. So far it appears to be very SM like, however there are discrepancies. One such discrepancy might be the coupling to two photons. The ATLAS collaboration measures the rate at which the Higgs decays to two photons to be above the predicted SM rate~\cite{Aad:2012tfa} while the CMS collaboration measures a defect in different productions~\cite{Chatrchyan:2013lba}. In the SM, the Higgs does not couple to photons at tree level, but only through loops of charged particles. Any model introducing new charged particles which couple to the Higgs has the potential to greatly affect the coupling to photons, since they enter at the same order. At this point, the observed rates are not compelling evidence of new physics, but leave the possibility open. As a natural solution to the hierarchy problem, models with supersymmetry (SUSY) are one of the preferred new physics scenarios, which also have the ability to alter this coupling from the corresponding SM value.

SUSY models also have several problems. For instance, models with general soft breaking terms pose issues in the flavor sector. Moreover, in minimal scenarios even generating the observed Higgs mass is not an easy task. Within the minimal supersymmetric standard model (MSSM) the Higgs boson mass is bounded at tree level by the mass of the $Z$ boson ($\sim$ 91 GeV). To raise this value to the observed 125.5 GeV, large quantum corrections are needed. Such large corrections can only come from top/stop loops and require either a large average mass of the stops or large stop mixing~\cite{Delgado:2012eu}. Large masses and/or mixings can lead to more fine tuning, introducing the so called little hierarchy problem. This problem can be relaxed significantly at the cost of giving up on minimality, e.g.~introducing extra particles that modify the tree-level prediction for the Higgs mass, allowing for naturally larger values of $m_h$. This kind of extensions would be of special interest if the new particles could also help to modify the MSSM rate of Higgs to diphoton decays.

One such class of models introduces an additional zero hypercharge ($Y$), $SU(2)_L$ triplet, chiral superfield to the MSSM~\cite{Agashe:2011ia, Delgado:2012sm, Delgado:2013zfa, Bandyopadhyay:2013lca}. This extra triplet is coupled to the Higgs fields, which has the effect of helping to raise the mass of the Higgs without the necessity of large stop masses or mixings. As an $SU(2)_L$ triplet with no hypercharge, the field contains singly electrically charged states. Due to the coupling with the Higgs, these additional charged states help to modify the MSSM rate of Higgs to diphoton decays. The additional triplet states couple with the Higgs, but not directly to quarks or gluons, so the production cross section for the triplet states at the LHC will be negligible. The presence of these extra states will only be inferred through their indirect effects on other processes (such as the $h\to\gamma \gamma$ ratio). Another way these extra states may be inferred is in stop decays, which will be the main focus of this paper.

Many different direct searches for supersymmetry, including searches for sfermions and gluinos, are being undertaken at the LHC~\cite{Chatrchyan:2013xna,Aad:2013ija, Chatrchyan:2013lya, Aad:2013wta}. In many of these searches a simplified model is used in which the stops decay either only to a top quark and a neutralino or only to a bottom quark and a chargino. This assumption is not always representative of the MSSM, but it allows for a search method. The addition of the triplet field introduces an extra chargino and neutralino. Despite the stop not having direct couplings to the triplet states, its decays can be modified due to mixing between the fermionic triplets and the Higgsinos. In this paper we demonstrate how this effect can introduce significant changes in the branching ratios of the stops, compared to the MSSM ones. These changes, of course, complicate the interpretation of direct searches analyses for stop pair production, as well as gluino pair production which cascades through stops.

The content of this paper goes as follows. In Section~\ref{section_TheModel} we will describe the setup of the model, including some considerations of the electroweak constraints on scalar triplets, as well as the parameters we will be using throughout this work. In Section~\ref{section_StopDecays} we examine the decays of the lightest stop in the triplet extension, and show how the branching ratios are affected by the fermion triplets, comparing the results with the MSSM branching ratios. In Section~\ref{Sec_LHC}, we do a quick LHC simulation and detector level analysis to exemplify how these changes can affect the direct searches of stop pairs. Finally, our conclusions are presented in Section~\ref{section_Conclusions}.


\section{The Model}
\label{section_TheModel}

We consider an extension of the MSSM with a $Y=0$ triplet chiral superfield. We follow the setup and notation of Ref.~\cite{Espinosa:1991wt}, and we refer to that reference for more details.  The triplet, $\Sigma$, can be described by its electrically charged and neutral components as 
\begin{equation}
\Sigma = \begin{pmatrix}
\xi^0/\sqrt{2} & -\xi^+_2 \\
\xi^-_1 & -\xi^0/\sqrt{2}
\end{pmatrix}.
\label{eqn_SigmaDefinition}
\end{equation}
The superpotential in the Higgs-$\Sigma$ sector is given by
\begin{equation}
W_{\mathrm{Higgs}\mbox{-}\Sigma} = \mu~\!H_1\!\cdot\! H_2+\lambda~\! H_1 \cdot \Sigma H_2+\frac{1}{2} \mu_{\Sigma}\tr \Sigma^2,
\label{eqn_SuperPotential}
\end{equation}
with $a\!\cdot\! b\equiv a^i\varepsilon_{ij}b^j$, $\varepsilon_{21}=-\varepsilon_{12}=1$, $\varepsilon_{11}=\varepsilon_{22}=0$. The soft breaking potential is also modified by
\begin{equation}
\Delta V_{\rm soft}=m_4^2 |\Sigma|^2+\left(\frac{1}{2}B_{\Sigma}\mu_\Sigma\tr\Sigma^2+A_\lambda \lambda~\! H_1 \cdot \Sigma H_2+\mathrm{h.c.}\right)~\!.
\label{eqn_soft}
\end{equation}
Compared with the MSSM, there are therefore five extra parameters:  the superpotential coupling $\lambda$, the supersymmetric mass $\mu_{\Sigma}$, the soft-breaking mass $m_4$, and the bilinear and trilinear soft-breaking parameters $B_{\Sigma}$ and $A_{\lambda}$. 

\subsection{Scalar triplet-Higgs sector}
\label{subsection_Scalar}

An expansion of the neutral scalar potential leads to
\begin{eqnarray}
V &=& m_1^2 |H_1^0|^2 +m_2^2 |H_2^0|^2 + m_4^2 |\xi^0|^2 + \frac{g^2+g^{\prime^2}}{8} (|H_2^0|^2-|H_1^0|^2)^2 \nonumber \\
 && +\left|\frac{\lambda}{\sqrt{2}}  H^0_2H^0_1-\mu_{\Sigma}\xi^0\right|^2 + \left| \frac{\lambda}{\sqrt{2}} H_1^0 \xi^0 - \mu H_1^0\right|^2 + \left|\frac{\lambda}{\sqrt{2}} H_2^0 \xi^0 - \mu H_2^0\right|^2 \nonumber \\
 &&+ (B_{\Sigma} \mu_{\Sigma} \xi^0 \xi^0-\frac{1}{\sqrt{2}}A_{\lambda} \lambda H^0_1 H^0_2 \xi^0-m_3^2 H_2^0H_1^0 + \mathrm{h.c.}).
\label{eqn_Vacuum}
\end{eqnarray}
Electroweak observables, specifically the $T$ parameter, constrain the vacuum expectation value (VEV) of $\xi^0$, $\langle\xi^0\rangle$, to be around the GeV scale or smaller~\cite{Delgado:2012sm}. A large mass for the scalar triplet lowers $\langle\xi^0\rangle$. Thus, the following hierarchy is used,  
\begin{equation}
\left|\frac{A_{\lambda}}{v}\right|, \left| \frac{\mu}{v} \right|, \left|\frac{\mu_{\Sigma}}{v}\right| \lesssim 10^{-2} \frac{m_{\Sigma}^2}{\lambda v^2},
\label{eqn_Hierarchy}
\end{equation}
where $m_{\Sigma}^2 = m_4^2 + \mu_{\Sigma}^2 + B_{\Sigma}\mu_{\Sigma}+\lambda^2 v^2/2$. For the rest of the paper, we will set the VEV of $\xi^0$ and the trilinear coupling $A_{\lambda}$ to zero. Under the conditions given by Eq.~(\ref{eqn_Hierarchy}) the scalar triplets decouple from the scalar doublets and the conditions needed to minimize the scalar potential are
\begin{eqnarray}
m_A^2 &=& m_1^2 + m_2^2 +2|\mu|^2 +\frac{\lambda^2}{2}v^2, \label{eqn_mA} \\
m_Z^2 &=& \frac{m_2^2 - m_1^2}{\cos (2\beta)} -m_A^2 +\frac{\lambda^2}{2}v^2, \label{eqn_mz}\\
m_3^2  &=&m_A^2\sin \beta \cos \beta,  \label{eqn_m3}
\end{eqnarray}
where $\tan\beta = v_2/v_1$ and $v=\sqrt{v_1^2+v_2^2} = 174$ GeV. Using the definitions $H^0_i = v_i +(h_i+i \chi_i)/\sqrt{2}$ and $x=\text{Re}\,\xi^0/\sqrt{2}$, the CP-even scalar mass squared terms are given at tree level by
\begin{equation}
\frac{1}{2}\begin{pmatrix} h_2 &  h_1\end{pmatrix} \mathcal{M}^2_{\text{CP-even}} \begin{pmatrix} h_2 \\ h_1 \end{pmatrix}
+\frac{1}{2}m_\Sigma^2 x^2, \nonumber 
\end{equation}
where
\begin{equation}
 \mathcal{M}^2_{\text{CP-even}} = 
\begin{pmatrix}
  m_A^2 \cos^2\beta + m_Z^2 \sin^2\beta & -(m_A^2 + m_Z^2 -v^2 \lambda^2) \cos \beta \sin \beta \\
 -(m_A^2 + m_Z^2 -v^2 \lambda^2) \cos \beta \sin \beta &m_A^2 \sin^2\beta + m_Z^2 \cos^2\beta 
\end{pmatrix}.
\label{eqn:EvenMassMatrix}
\end{equation}

We will examine what happens for a range of values for the triplet fermion mass parameter, $\mu_{\Sigma}$. At the level of the one-loop effective potential, the CP-even neutral Higgs mixing matrix is changed by the triplet as
\begin{eqnarray}
\Delta_{\Sigma}\mathcal{M}^2_{11} &=& \frac{5\lambda^4}{32\pi^2} \log\left(\frac{m^2_{\Sigma}}{\mu_{\Sigma}^2}\right) v^2 \sin^2\beta, \nonumber \\
\Delta_{\Sigma}\mathcal{M}^2_{22} &=& \frac{5\lambda^4}{32\pi^2} \log\left(\frac{m^2_{\Sigma}}{\mu_{\Sigma}^2}\right) v^2 \cos^2\beta, \nonumber \\
\Delta_{\Sigma}\mathcal{M}^2_{12} &=& \frac{\lambda^4}{32\pi^2} \log\left(\frac{m^2_{\Sigma}}{\mu_{\Sigma}^2}\right) v^2 \sin\beta\cos\beta .
\label{eqn_TripMassEffects}
\end{eqnarray}
This shows that the mass of the Higgs depends on the mass parameter of the fermion triplets $\mu_{\Sigma}$ through radiative corrections. The reader is referred to Ref.~\cite{Delgado:2013zfa} for more information on the one-loop effective potential effects of the stop and triplet sectors on the mass and mixings of the Higgs.

\begin{figure}[t]
        \centering
        \begin{subfigure}[b]{0.45\textwidth}
                \centering
                \includegraphics[width=\textwidth]{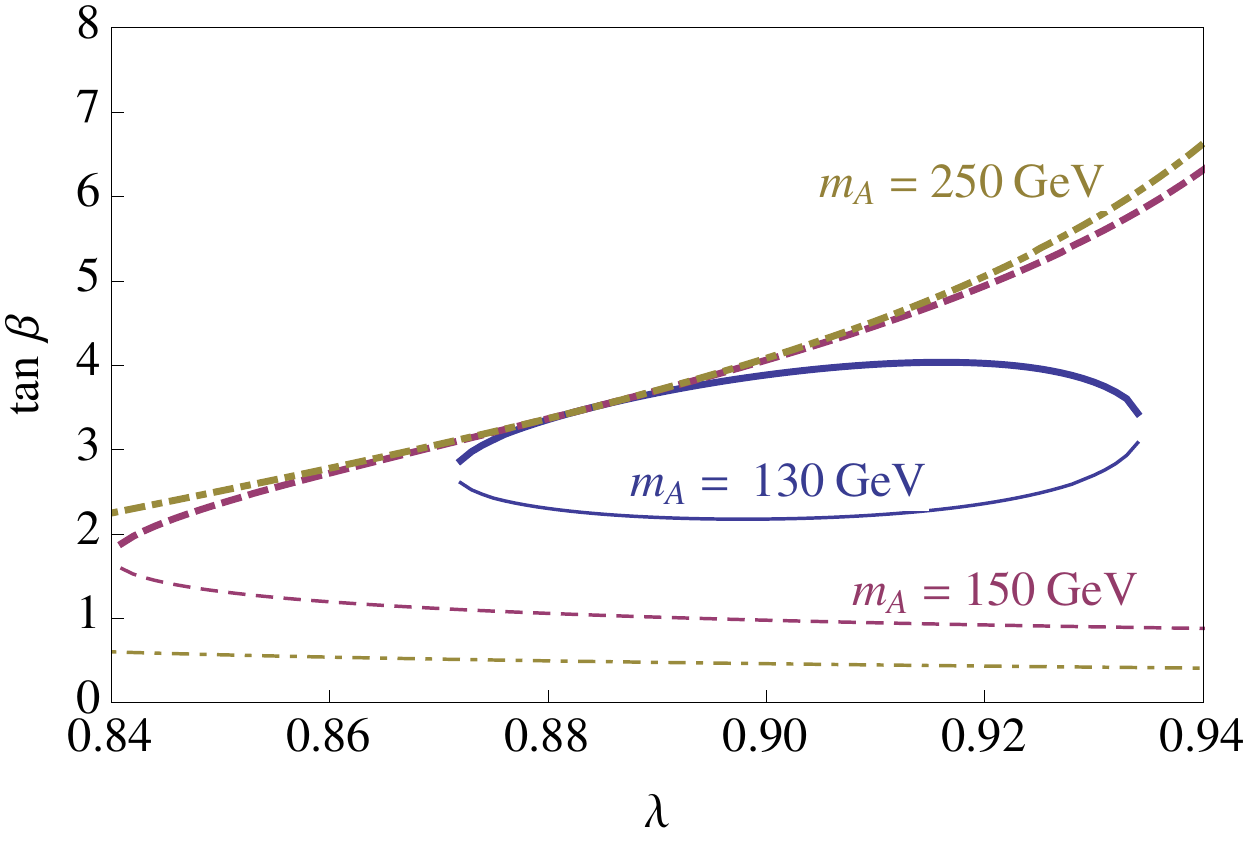}
                \caption{}
                \label{fig:CritPoint200}
        \end{subfigure}%
        \qquad
        \begin{subfigure}[b]{0.45\textwidth}
                \centering
                \includegraphics[width=\textwidth]{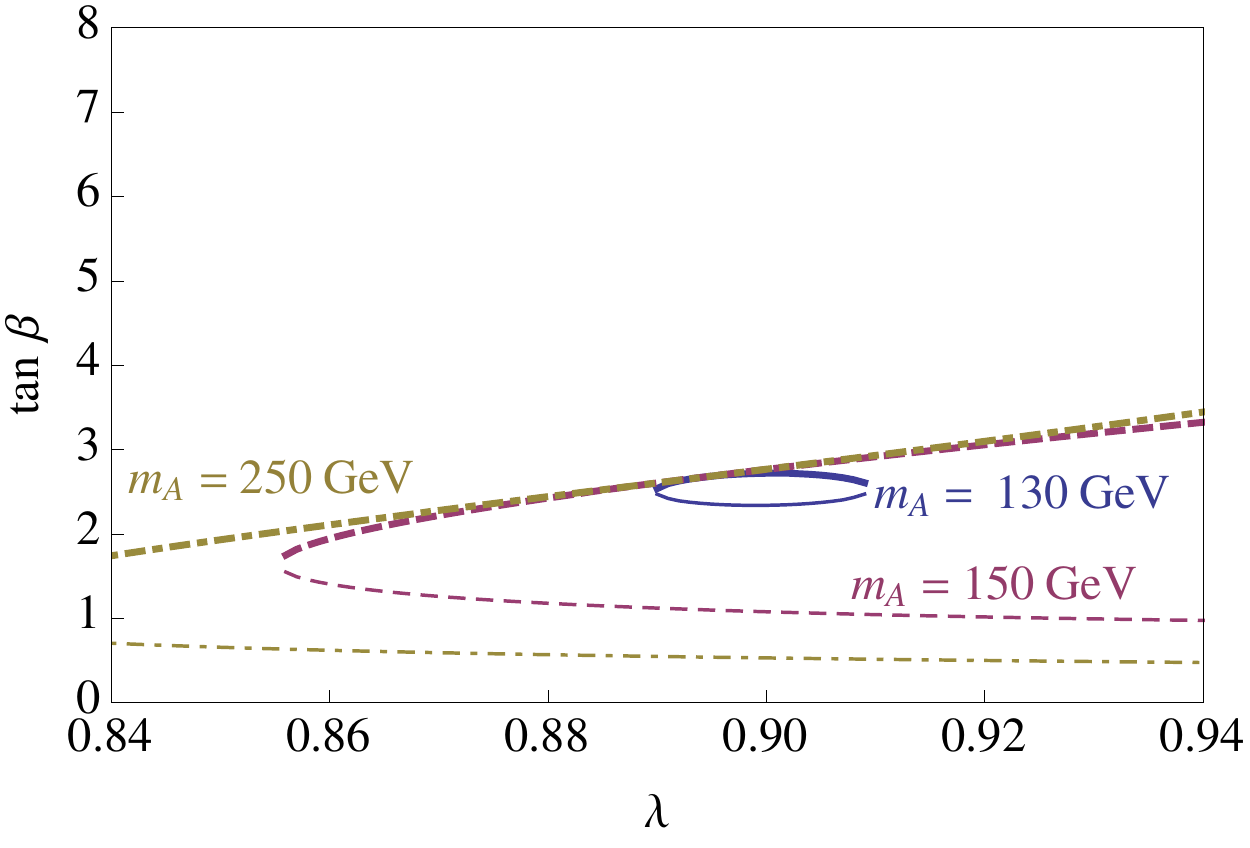}
                \caption{}
                \label{fig:CritPoint800}
        \end{subfigure}
        \caption{\it Regions where the Higgs mass is 125.5 GeV when $\mu_{\Sigma} = 200$ GeV (\subref{fig:CritPoint200}) and  $\mu_{\Sigma} = 800$ GeV  (\subref{fig:CritPoint800}) for different values of $m_A$ and $m_{U^c_3} = m_{Q_3} = 700$ GeV.}
        \label{fig_CriticalPoints}
\end{figure}

One interesting result in Ref.~\cite{Delgado:2013zfa} is the existence of critical points $(\lambda, \tan\beta)=(\lambda_C, \tan\beta_C)$ where the decoupling or alignment limit occurs for much smaller values of $m_A$ than in the MSSM~\footnote{For a recent analysis of alignment in different models see Ref.~\cite{Carena:2013ooa}.}. The curves corresponding to a fixed value of the Higgs mass in $(\lambda, \tan\beta)$ space do intersect for different fixed values of $m_A$. Examples of this are shown in Fig.~\ref{fig_CriticalPoints} where the Higgs mass is fixed at 125.5 GeV. At this critical point, the value of the rotation of the VEVs ($\beta$) and the rotation angle ($\alpha$) for the mass eigenstates of the two Higgs doublets align such that $\sin(\beta - \alpha) = 1$. In this case, the couplings of the lightest Higgs boson to vector fields are the same as in the SM. The region around the critical point is similar, with $\sin (\beta-\alpha) \sim 1$.
As the value of $\mu_{\Sigma}$ increases, the critical value of $\tan\beta$ decreases, while raising the minimum value of $m_A$ which can yield a correct Higgs mass, as would be expected in the decoupling limit of the MSSM.  However, although it is clear that one strictly recovers the MSSM in the limit $\mu_{\Sigma}\to\infty$, for large values $\mu_\Sigma\gtrsim m_{\tilde{t}_i}$ the triplet fermions will not affect the decay of stops but still (as the triplet is not integrated out supersymmetrically) the coupling $\lambda$ in the superpotential provides a tree-level contribution to the Higgs mass which makes the model phenomenologically viable, as can be seen in the right panel of Fig.~\ref{fig_CriticalPoints}~\footnote{Note that in the strict MSSM limit, using  the specific values $m_{Q_3} = m_{U^c_3} =700$ GeV and $A_t = 0$ will not generate the observed mass of the Higgs.}. To simplify the notation we will call such models MSSM-like.

\begin{figure}[t]
        \centering
        \begin{subfigure}[b]{0.45\textwidth}
                \centering
                \includegraphics[width=\textwidth]{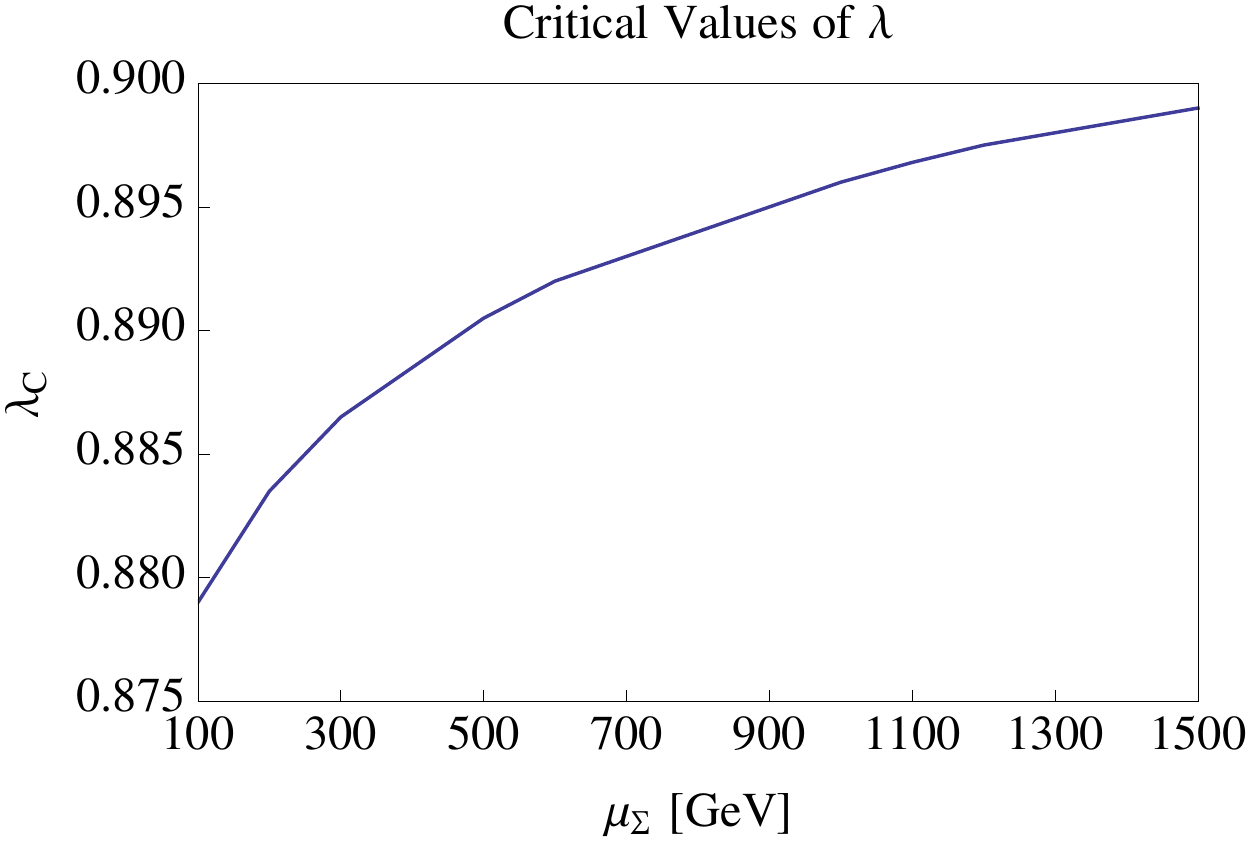}
                \caption{}
                \label{fig:LambdaCrit}
        \end{subfigure}%
        \qquad
        \begin{subfigure}[b]{0.45\textwidth}
                \centering
                \includegraphics[width=\textwidth]{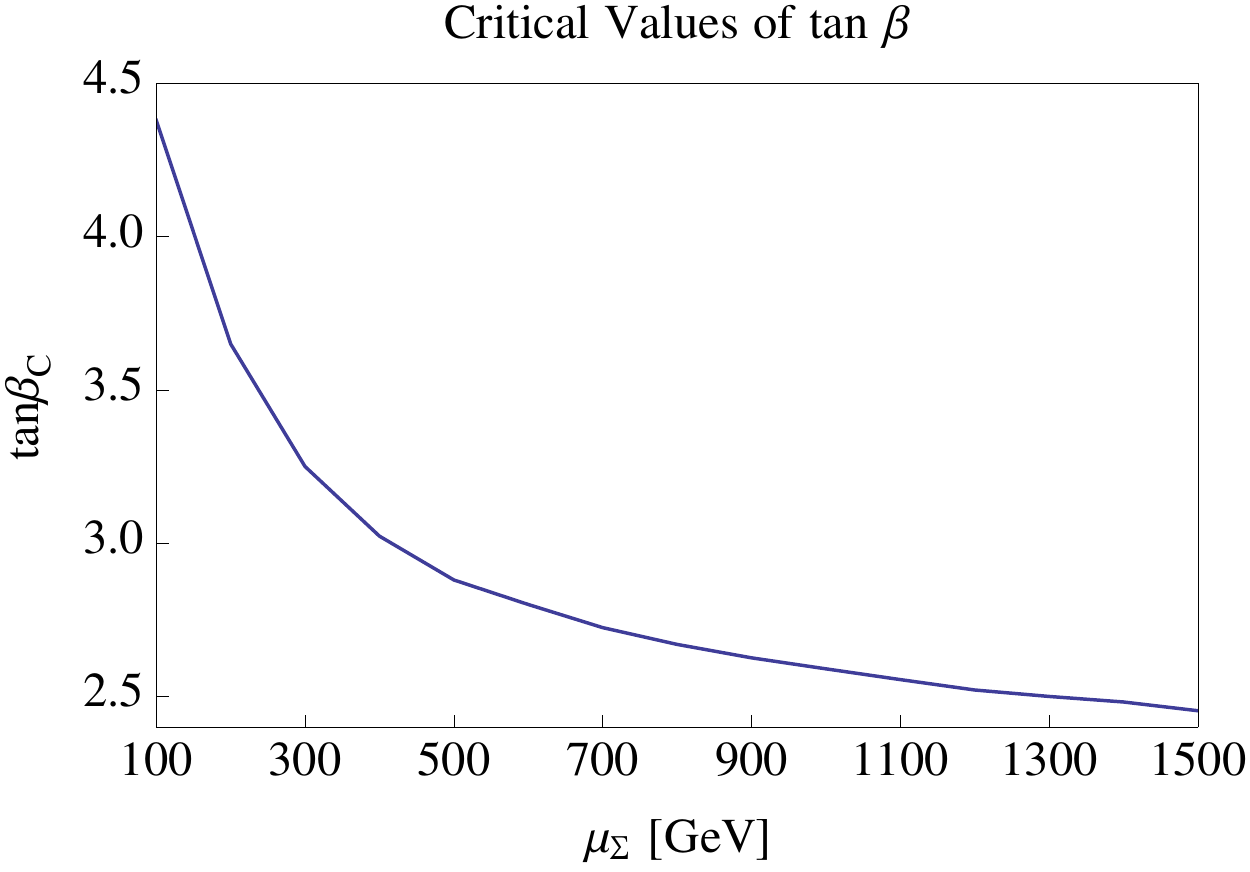}
                \caption{}
                \label{fig:TanBeta}
        \end{subfigure}
        \caption{The critical values of $\lambda$ (\subref{fig:LambdaCrit}) and $\tan\beta$ (\subref{fig:TanBeta}) as a function of $\mu_{\Sigma}$. }
        \label{fig_CriticalValues}
\end{figure}

In Fig.~\ref{fig_CriticalValues} the critical values of $\lambda$ and $\tan\beta$ are plotted against $\mu_{\Sigma}$, for $m_{Q_3} = m_{U^c_3} =700$ GeV and $A_t = 0$. In the following we will always use the critical values $(\lambda_C, \tan\beta_C)$ for the specific values of $\mu_{\Sigma}$, $m_{Q_3}$ and $m_{U_3^c}$ being analyzed. For simplicity, we will keep $A_t=0$ to reduce the number of free parameters. Keeping a small value of $A_t$ can be justified by naturalness arguments (note that in this model neither the stop masses or mixing have to be large). On the other hand, if all $A$-terms have a common origin, since $A_\lambda$ has to be kept small to avoid conflict with electroweak precision data constraints, $A_t$ will also be expected to be small. This occurs naturally, for instance, in scenarios with gauge mediation of SUSY breaking.

\subsection{Chargino and Neutralino sectors}
\label{subsection_Ino}

The addition of the triplet chiral superfield not only helps raise the mass of the Higgs, but it also adds an extra chargino and neutralino. In the (enlarged) gauge-eigenstate basis, $\psi^0=\begin{pmatrix} \widetilde{B} & \widetilde{W}^0 & \widetilde{H}_1^0 & \widetilde{H}_2^0 & \widetilde{\xi}^0\end{pmatrix}$, $\psi^+=\begin{pmatrix}\widetilde{W}^+ & \widetilde{H}^+_2 & \widetilde{\xi}^+_2\end{pmatrix}$, $\psi^-=\begin{pmatrix} \widetilde{W}^- & \widetilde{H}^-_1 & \widetilde{\xi}^-_1\end{pmatrix}$, the corresponding mass terms are given by:
\begin{equation}
\Delta {\cal L}=-\frac{1}{2}\begin{pmatrix}(\psi^+)^T&(\psi^-)^T\end{pmatrix}\left(\begin{matrix}0& \mathcal{M}_{\widetilde{C}}^T\\\mathcal{M}_{\widetilde{C}}&0\end{matrix}\right)\begin{pmatrix}\psi^+\\ \psi^-\end{pmatrix}
-\frac{1}{2}(\psi^0)^T\mathcal{M}_{\widetilde{N}}\psi^0 +\mathrm{h.c.},
\end{equation}
with mass matrices given by:

\begin{equation}
\mathcal{M}_{\widetilde{C}} = \begin{pmatrix} M_2 & gv\; \sin \beta & 0 \\ gv \cos \beta & \mu & -\lambda v \sin\beta \\ 0 & -\lambda v \cos \beta & -\mu_{\Sigma} \end{pmatrix},
\label{eqn_CharginoMatrix}
\end{equation}
\begin{eqnarray}
\mathcal{M}_{\widetilde{N}} &=& 
\begin{pmatrix}
M_1 & 0 & -\cos \beta s_W m_Z & \sin \beta s_W m_Z & 0 \\
0 & M_2 & \cos \beta c_W m_Z & -\sin \beta c_W m_Z & 0 \\
-\cos \beta s_W m_Z & \cos \beta c_W m_Z & 0 & -\mu & \frac{\lambda}{\sqrt{2}} v \sin \beta \\
\sin \beta s_W m_Z & -\sin \beta c_W m_Z & -\mu & 0 & \frac{\lambda}{\sqrt{2}} v \cos \beta \\
0 & 0 &\frac{\lambda}{\sqrt{2}} v \sin \beta & \frac{\lambda}{\sqrt{2}} v \cos \beta & \mu_{\Sigma}
\end{pmatrix},\nonumber
\end{eqnarray}
where $s_W\equiv \sin{\theta_W}$, $c_W\equiv \cos{\theta_W}$. 

The rest of this paper will focus on finding regions of parameter space in which the presence of the extra neutralino and chargino could be inferred. Searches for supersymmetry often focus on colored particles due to the strength of the strong coupling. In many top-down SUSY models, the lightest stop is the lightest sfermion because of the effect on the running of soft masses from the top Yukawa coupling. We choose a region of parameter space where the stop is the lightest sfermion. Assuming a large enough value for the gluino mass, the stop can only decay to a chargino or a neutralino and SM particles. 

\begin{figure}[t]
        \centering
        \begin{subfigure}[b]{0.45\textwidth}
                \centering
                \includegraphics[width=\textwidth]{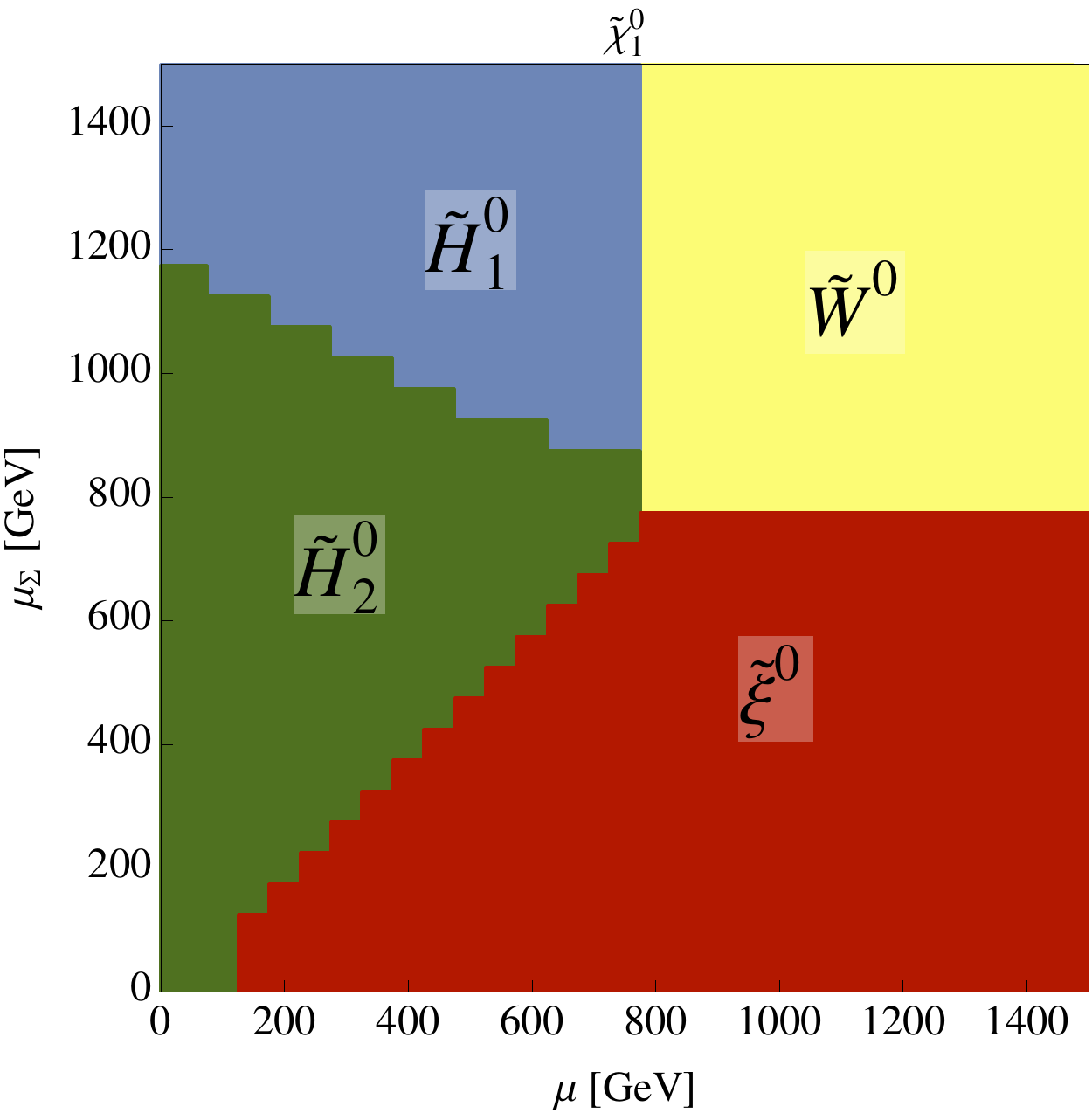}
                \caption{}
                \label{fig_NeuDis1}
        \end{subfigure}%
        \qquad
        \begin{subfigure}[b]{0.45\textwidth}
                \centering
                \includegraphics[width=\textwidth]{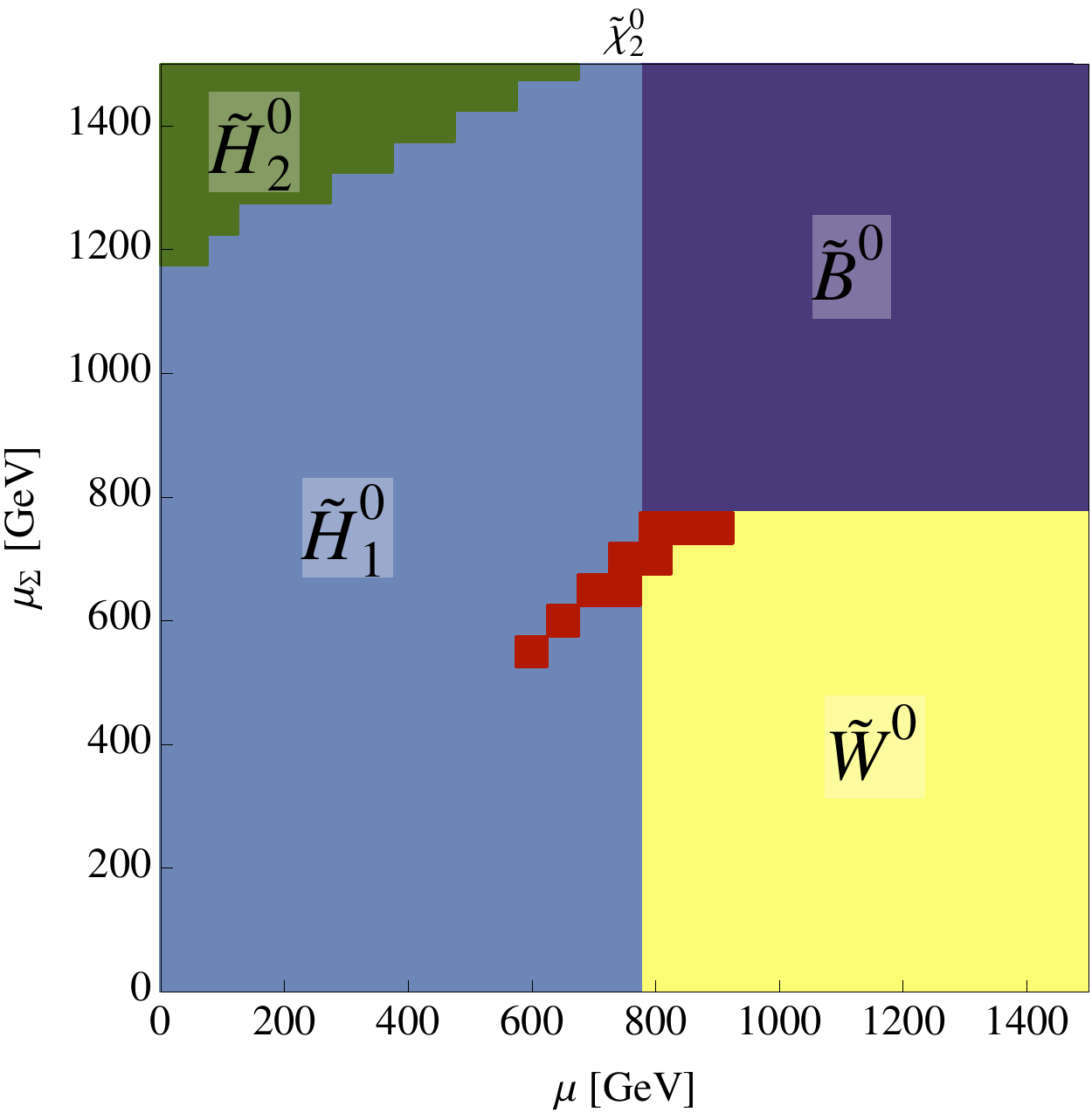}
                \caption{}
                \label{fig_NeuDis2}
        \end{subfigure}
                
          \begin{subfigure}[b]{0.45\textwidth}
                \centering
                \includegraphics[width=\textwidth]{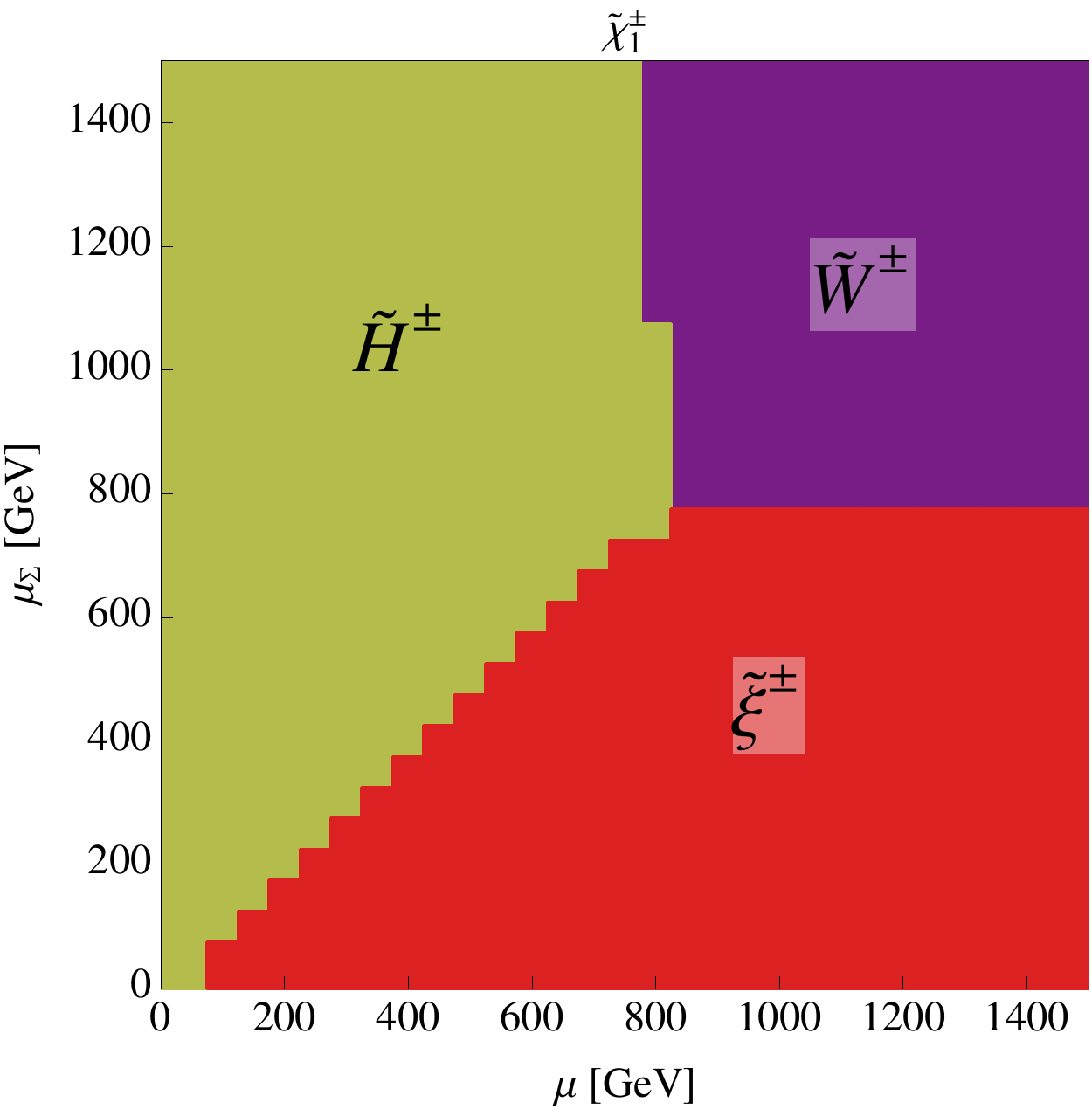}
                \caption{}
                \label{fig_CharDist1}
        \end{subfigure}%
        \qquad
        \begin{subfigure}[b]{0.45\textwidth}
                \centering
                \includegraphics[width=\textwidth]{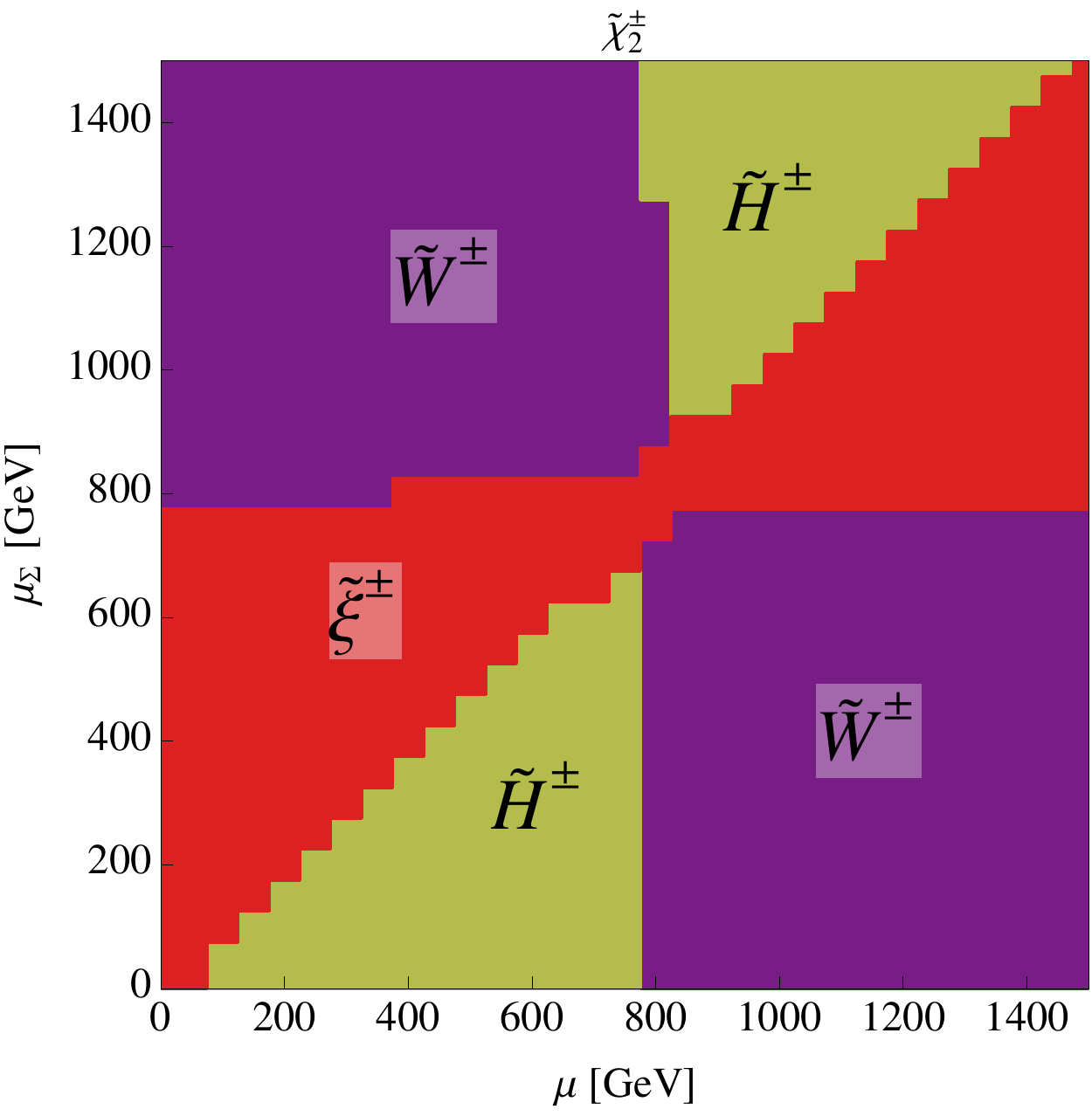}
                \caption{}
                \label{fig_CharDist2}
        \end{subfigure}
        \caption{The interaction eigenstate which is the largest portion of (\subref{fig_NeuDis1}) $\widetilde{\chi}^0_1$ [lightest neutralino], (\subref{fig_NeuDis2}) $\widetilde{\chi}^0_2$ [next-to-lightest neutralino], (\subref{fig_CharDist1}) $\widetilde{\chi}^{\pm}_1$ [lightest chargino], and (\subref{fig_CharDist2}) $\widetilde{\chi}^{\pm}_2$ [next-to-lightest chargino].}
        \label{fig_Distributions}
\end{figure}

Regarding the stop sector, as explained in the previous section, we will fix $A_t=0$. For the third generation of squark soft masses, for illustrational purposes we will consider the following three benchmark points, which represent the cases of similar stop masses and large stop mixing ({\it Point A}), the lightest stop being mostly right-handed ({\it Point B}), or being mostly left-handed ({\it Point C}):
\begin{eqnarray}
\mbox{Point A:}&m_{Q_3}=m_{U_3^c}=700~\mathrm{GeV},\nonumber\\
\mbox{Point B:}&m_{Q_3}=1100~\mathrm{GeV},~~m_{U_3^c}=700~\mathrm{GeV},\label{points}\\
\mbox{Point C:}&m_{Q_3}=700~\mathrm{GeV},~~m_{U_3^c}=1100~\mathrm{GeV}.\nonumber
\end{eqnarray}
The fermionic triplets do not couple to the stops, so they only affect the stop decays through mixing. Examination of the mass matrices in Eq.~(\ref{eqn_CharginoMatrix}) reveals that the triplet fermions can only mix with the Higgsino states.\footnote{Mixing between the triplet charged fermions and Winos is allowed in principle, via the VEV of $\xi^0$. But, as explained above, this is strongly constrained by the electroweak precision observables and we neglect it.} Therefore, the triplet effects on the stop phenomenology will only manifest when the stop decays into states that are mostly fermion triplet or Higgsino like. Decays into a state which is mostly fermion triplet like will be suppressed. Similarly the strongest decays will be to states which are mostly  $\widetilde{H}_2$-like, due to the top Yukawa coupling. Therefore, to see the effect of the triplets, we have set the Bino and Wino masses, $M_1$ and $M_2$, above the lightest stop, and we scan only over the  supersymmetric mass of the fermion triplet, $\mu_{\Sigma}$, and the Higgsino mass parameter, $\mu$. Likewise, as mentioned above, we will also choose a value for the gluino mass such that decays into these states are not allowed. We then take the following values for the gaugino masses,
\begin{equation}
M_1=M_2=800~\mathrm{GeV}~~\mbox{and}~~M_3=1000~\mathrm{GeV}.
\end{equation}
For these values of the parameters we will always work in the critical point, $\tan\beta=\tan\beta_C$ and $\lambda=\lambda_C$, which allows to reproduce the Higss mass for small $m_A$ values, which we set as $m_A=130$ GeV. In Fig.~\ref{fig_Distributions} the interaction eigenstates that account for the largest component of the lightest and next-to-lightest neutralinos and charginos are displayed over a range of $\mu$ and $\mu_{\Sigma}$ for {\it point A}. As an example, for the lightest neutralino, this was calculated using the unitary matrix {\bf N} which diagonalizes $\mathcal{M}_{\widetilde{N}}$. Defining the lightest neutralino by $\widetilde{\chi}_1^0= {\bf N}_{1j}\psi^0_{j}$, the plot in the upper left panel of Fig.~\ref{fig_Distributions} shows the component $\psi^0_{j}$ with the maximum value of ${\bf N}_{1j}^2$. 

To better illustrate the effects that a relatively light fermion triplet has on the stops decays, in the next section we will compare the results in the triplet extension to that of the MSSM-like models (i.e.~for large values of $\mu_\Sigma$) using the same parameter selection. In particular, we will use for $\tan{\beta}$ the critical value obtained for the corresponding light triplet extension scenario. In those MSSM-like scenarios the decays of the stops are going to be the same ones of the MSSM since there are no extra light states.

\section{Stop Decays}
\label{section_StopDecays}

\begin{figure}[h]
	\centering
	\includegraphics[width=0.45 \textwidth]{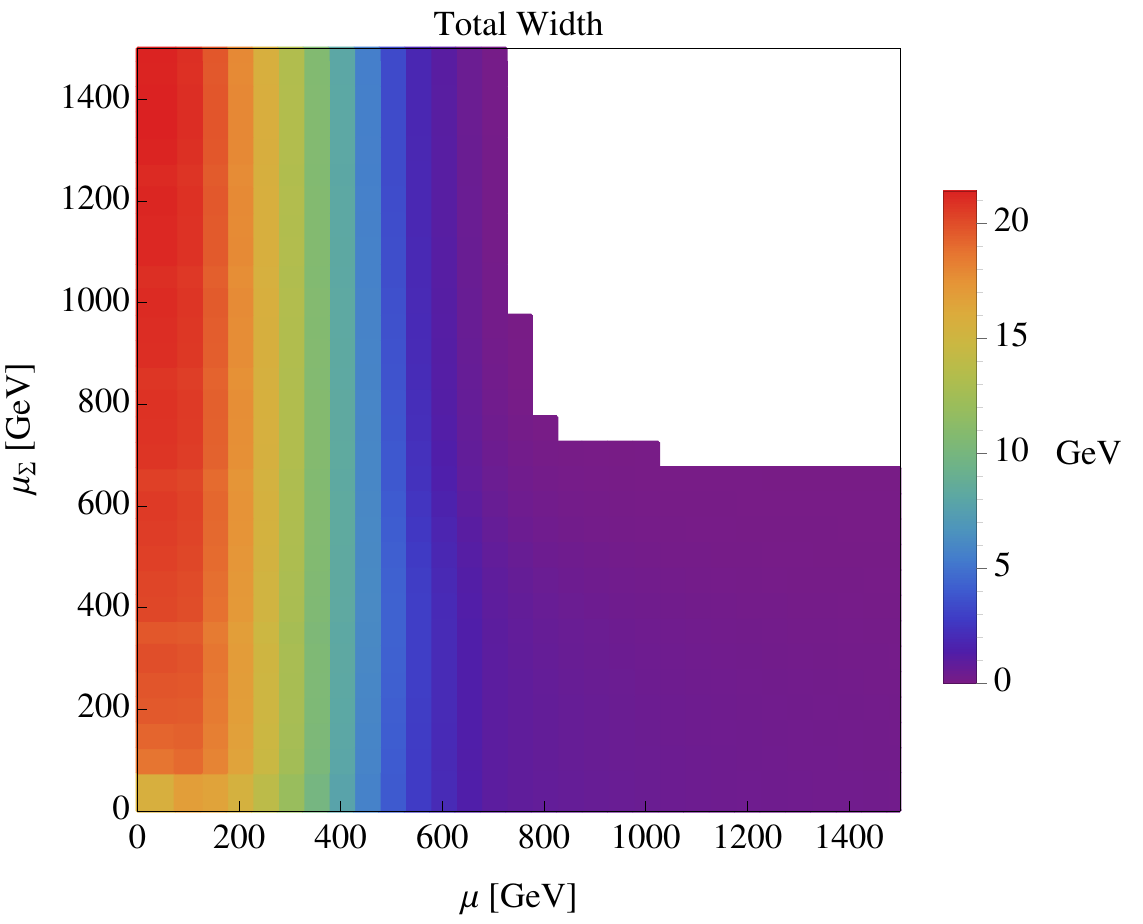}
	\caption{Total width of the lightest stop. Note that the width does not vary with $\mu_{\Sigma}$ other than the upper right corner of parameter space. In this area the stop has no open decay channels.}
	\label{fig_TotalWidth}
\end{figure}

\begin{figure}[t]
        \centering
        \begin{subfigure}[b]{0.48\textwidth}
                \centering
                \includegraphics[width=\textwidth]{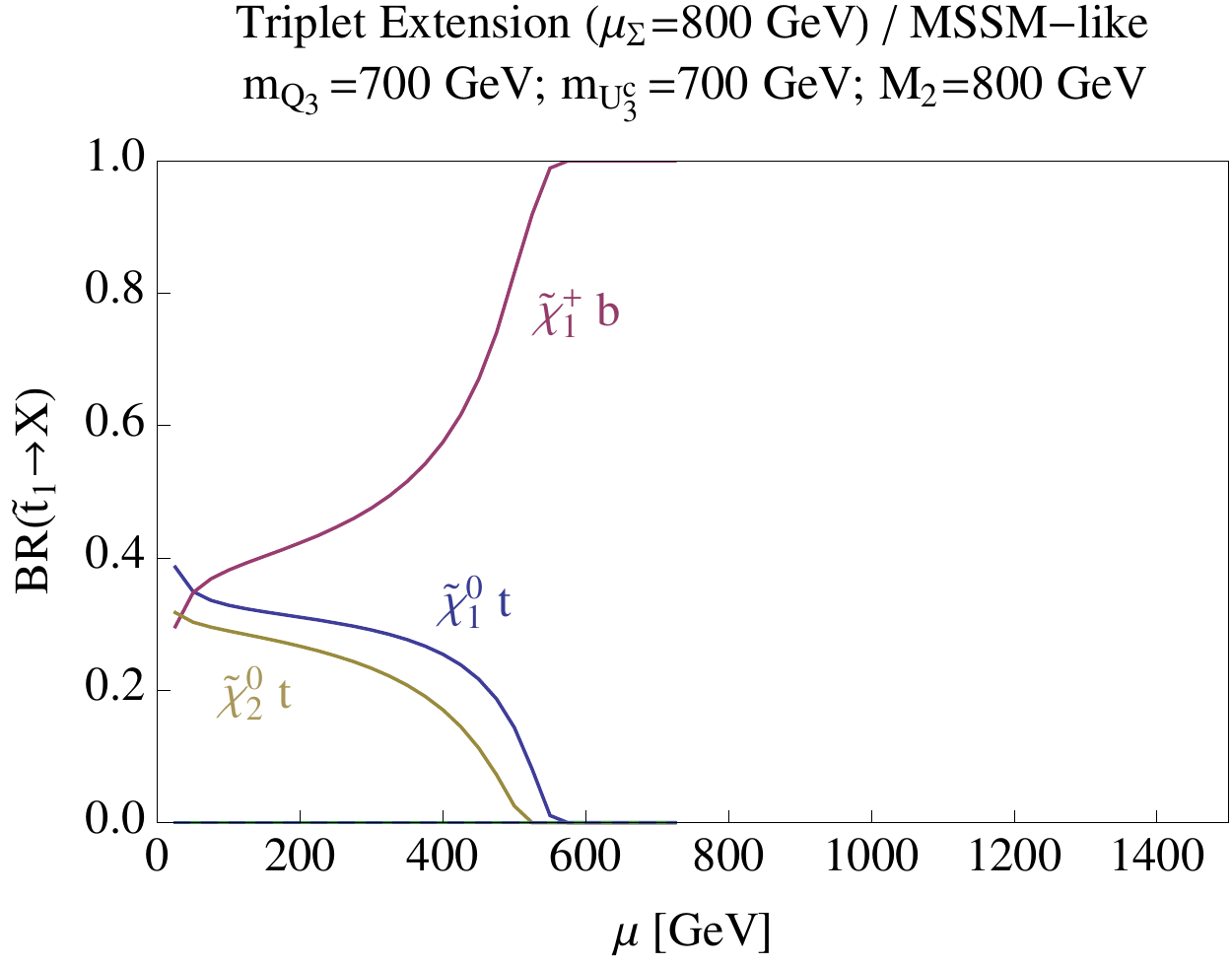}
                \caption{}
                \label{fig_pA_H}
        \end{subfigure}%
        \quad
        \centering
         \begin{subfigure}[b]{0.48\textwidth}
                \centering
                \includegraphics[width=\textwidth]{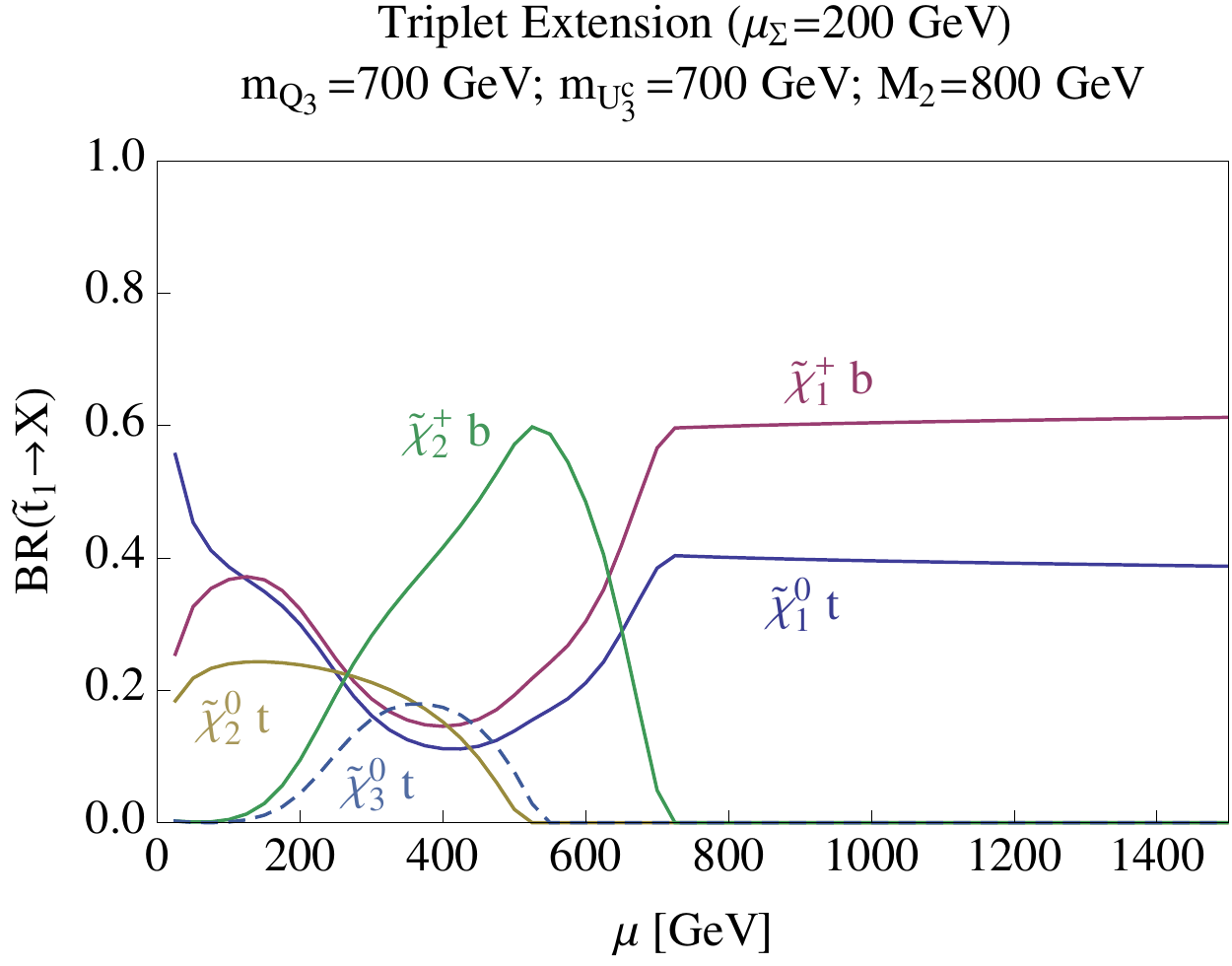}
                \caption{}
                \label{fig_pA_T}
        \end{subfigure}
        \quad
        \begin{subfigure}[b]{0.48\textwidth}
                \centering
                \includegraphics[width=\textwidth]{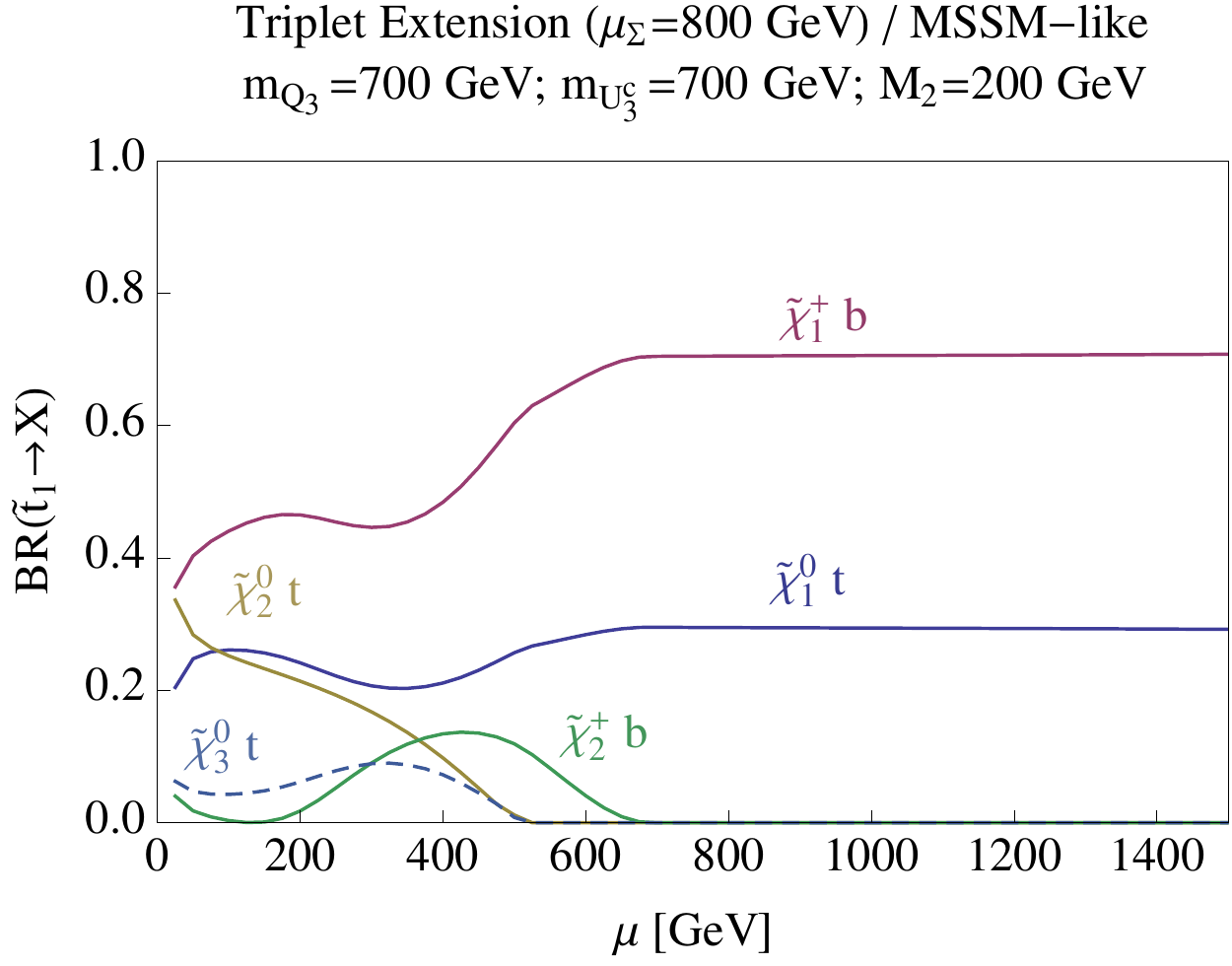}
                \caption{}
                \label{fig_pA_W}
        \end{subfigure}%
        \caption{Comparison of the branching ratios (BR) of the lightest stop, as a function of $\mu$, for {\it point A}: $m_{Q_3}=m_{U_3^c}=700$ GeV. (\subref{fig_pA_H}) In the triplet extension for $\mu_{\Sigma} = 800$ GeV (at the critical point). This reproduces the MSSM with $\tan\beta$ set to the same critical value of 2.67. (\subref{fig_pA_T}) In the triplet extension with $\mu_{\Sigma} = 200$ GeV.  In this case the critical values are $(\tan\beta_C,\lambda_C)=(3.50,0.884)$. (\subref{fig_pA_W}) The same in the MSSM-like case, with $\tan\beta=3.50$ and $M_2 = 200$ GeV so the stop also has a light triplet state available for decays. \label{fig_Point_A}}
\end{figure}

Because of the absence of any direct coupling between the triplet and the sfermions, the total width of the stop is mostly independent on the value of $\mu_{\Sigma}$. This can be seen in Fig.~\ref{fig_TotalWidth}, where we plot the total width of the lightest stop for {\it point A}.
From here on, the branching ratios for the different decays of the lightest stop will be calculated in the triplet extension and in the MSSM-like scenarios. First of all, as the parameter $\mu_\Sigma$ gets large, and the fermion triplet becomes decoupled, the results should approach the MSSM limit.
The decoupling can be seen in Fig.~\ref{fig_pA_H}, where we plot, as a function of $\mu$, the branching ratios of the lightest stop in the triplet extension at {\it point A} for $\mu_{\Sigma} = 800$ GeV. The critical value of $\tan\beta$ in this case is 2.67.
This exactly reproduces the branching ratios of the MSSM for the same value of $\tan\beta$, with the decay $\widetilde{t}_1 \rightarrow \widetilde{\chi}^{+}_1 \,b$ dominating at large values of $\mu$.

When the fermion triplets are not decoupled, an extra decay mode is opened for the stop. In Fig.~\ref{fig_pA_T} we plot the branching ratios for {\it point A} for $\mu_{\Sigma} = 200$ GeV, which gives $\tan\beta_C = 3.50$. The first feature this figure shows is how a stop decay is allowed for $\mu \ge m_{\widetilde{t_1}}$, which is not true for the MSSM or the large $\mu_{\Sigma}$ limit. The results for the light fermion triplet case can be compared to the MSSM-like model with $\mu_\Sigma=800$ GeV and a small value of $M_2$, so the stop has a triplet state to decay into for both cases (the states $\tilde\xi^i$ for the triplet extension and $\widetilde W^i$ for the MSSM-like case). The results in this MSSM-like scenario can be seen in Fig. \ref{fig_pA_W}, where we have considered $M_2=200$ GeV. 
To make the comparison faithful we have fixed for $\tan\beta$ the same critical value as in Fig.~\ref{fig_pA_T}, i.e.~$\tan\beta=3.50$. This means that in order to reproduce the Higgs mass we need, from the right panel of Fig.~\ref{fig_CriticalPoints}, $\lambda\simeq 0.94$ and large values of $m_A$, the decoupling regime of the MSSM. 
This extends the decays to regions where the Higgsinos are heavier than the stop. By comparing Figs.~\ref{fig_pA_T} and \ref{fig_pA_W} we observe that the most significant difference in the triplet extension is an enhancement of the decays to the second chargino state for $\mu_\Sigma<\mu\lesssim m_{\tilde{t}_1}$, being the dominant channel in most of this region. Indeed for $\mu>200\units{GeV}$ the second chargino becomes Higgsino like in both cases, while the lightest neutralino/chargino states become mostly triplet (Wino) in the triplet extension (MSSM-like). Thus, in the triplet scenario the (small) mixing with the Higgsinos suppresses the $\tilde{\chi}^{0,\pm}_1$ channels, compared to the MSSM-like case where they occur through $SU(2)_L$ gauge couplings, and in turn enhances the branching ratio to the $\tilde{\chi}^\pm_2$ channel. 

\begin{figure}[t]
        \centering
        \begin{subfigure}[b]{0.48\textwidth}
                \centering
                \includegraphics[width=\textwidth]{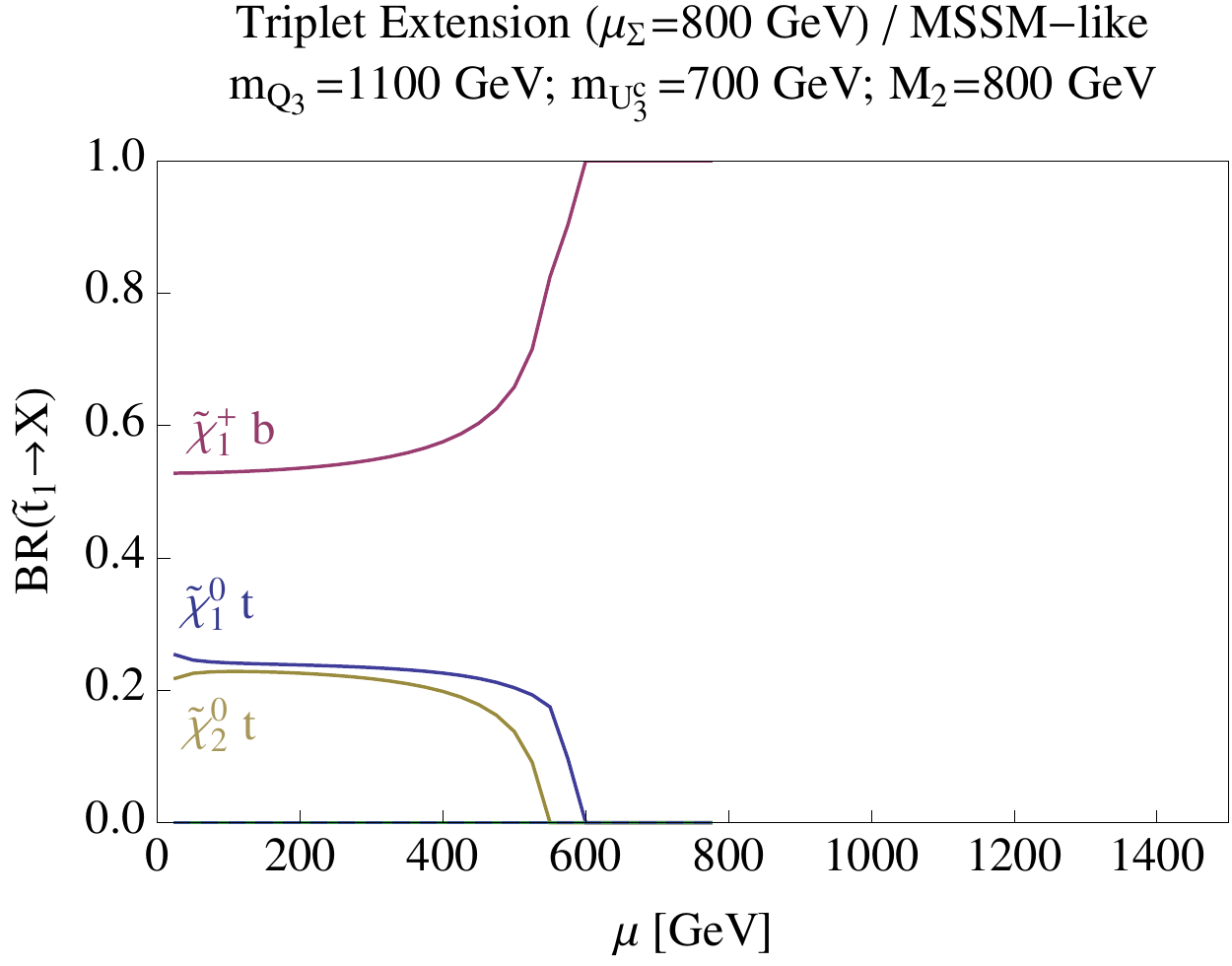}
                \caption{}
                \label{fig_pB_H}
        \end{subfigure}%
        \quad
        \centering
         \begin{subfigure}[b]{0.48\textwidth}
                \centering
                \includegraphics[width=\textwidth]{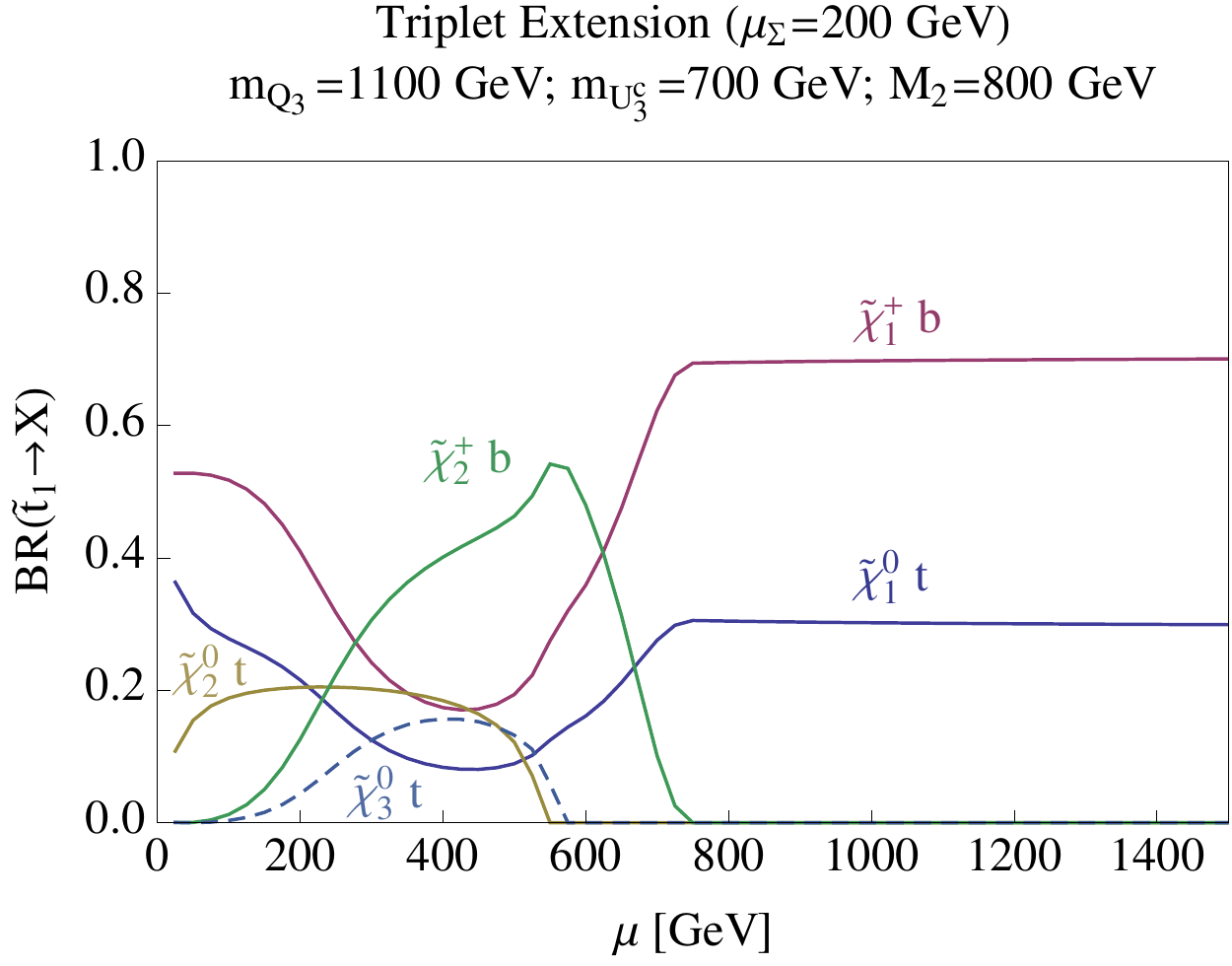}
                \caption{}
                \label{fig_pB_T}
        \end{subfigure}
        \quad
         \centering
         \begin{subfigure}[b]{0.48\textwidth}
                \centering
                \includegraphics[width=\textwidth]{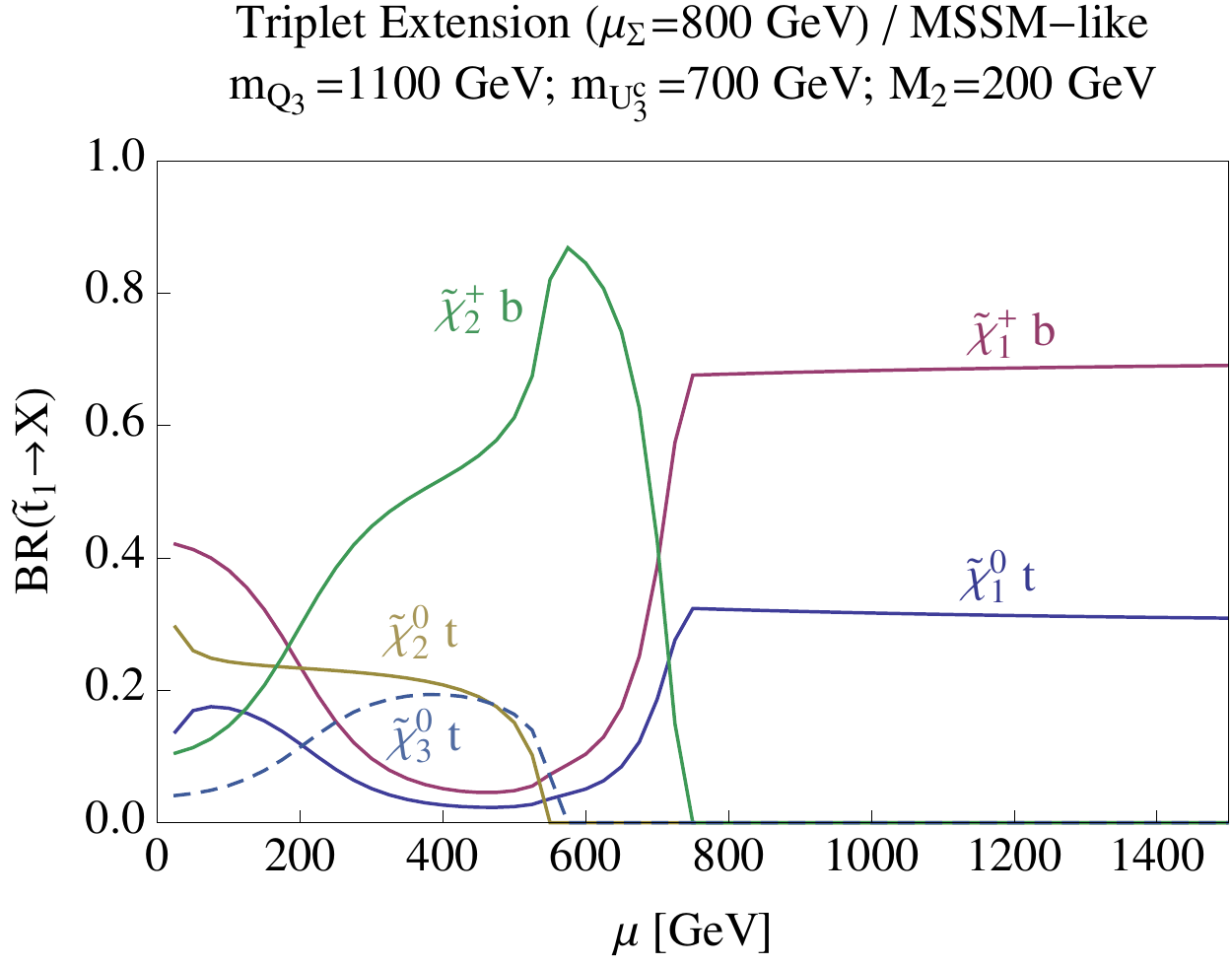}
                \caption{}
                \label{fig_pB_W}
        \end{subfigure}
        \caption{Comparison of the branching ratios of the lightest stop, as a function of $\mu$, for {\it point B}: $m_{Q_3}=1100$ GeV, $m_{U_3^c}=700$ GeV. (\subref{fig_pB_H}) In the triplet extension for $\mu_{\Sigma} = 800$ GeV (at the critical point) or, equivalently, the MSSM with $\tan\beta=3.48$. (\subref{fig_pB_T}) In the triplet extension with $\mu_{\Sigma} = 200$ GeV.  In this case the critical values are $(\tan\beta_C,\lambda_C)=(5.99,0.876)$. (\subref{fig_pB_W}) The same in the MSSM-like case, with $\tan\beta=5.99$ and $M_2 = 200$ GeV so the stop also has a light triplet state available for decays. \label{fig_Point_B}}
\end{figure}

\begin{figure}[t]
        \centering
        \begin{subfigure}[b]{0.48\textwidth}
                \centering
                \includegraphics[width=\textwidth]{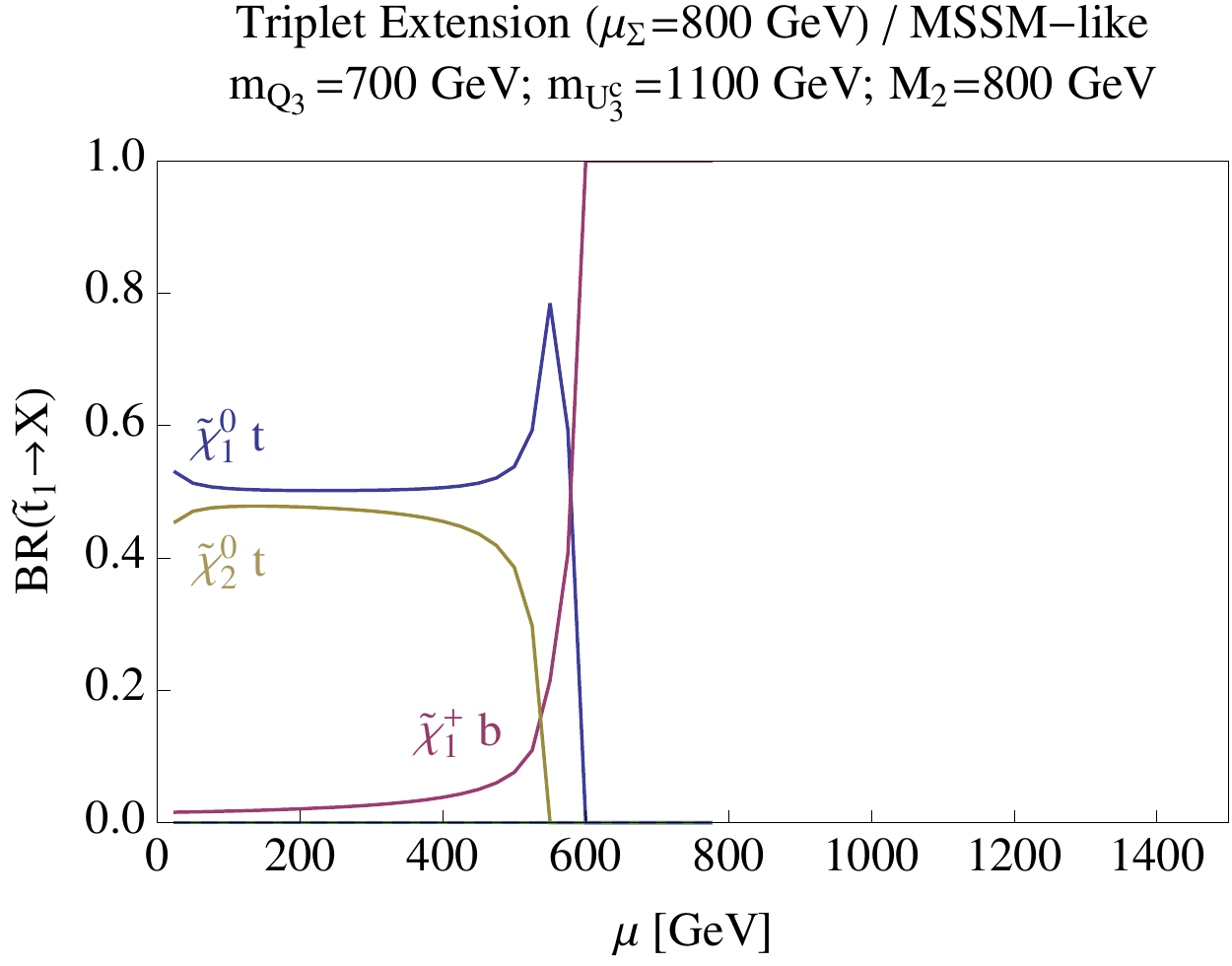}
                \caption{}
                \label{fig_pC_H}
        \end{subfigure}%
        \quad
        \centering
         \begin{subfigure}[b]{0.48\textwidth}
                \centering
                \includegraphics[width=\textwidth]{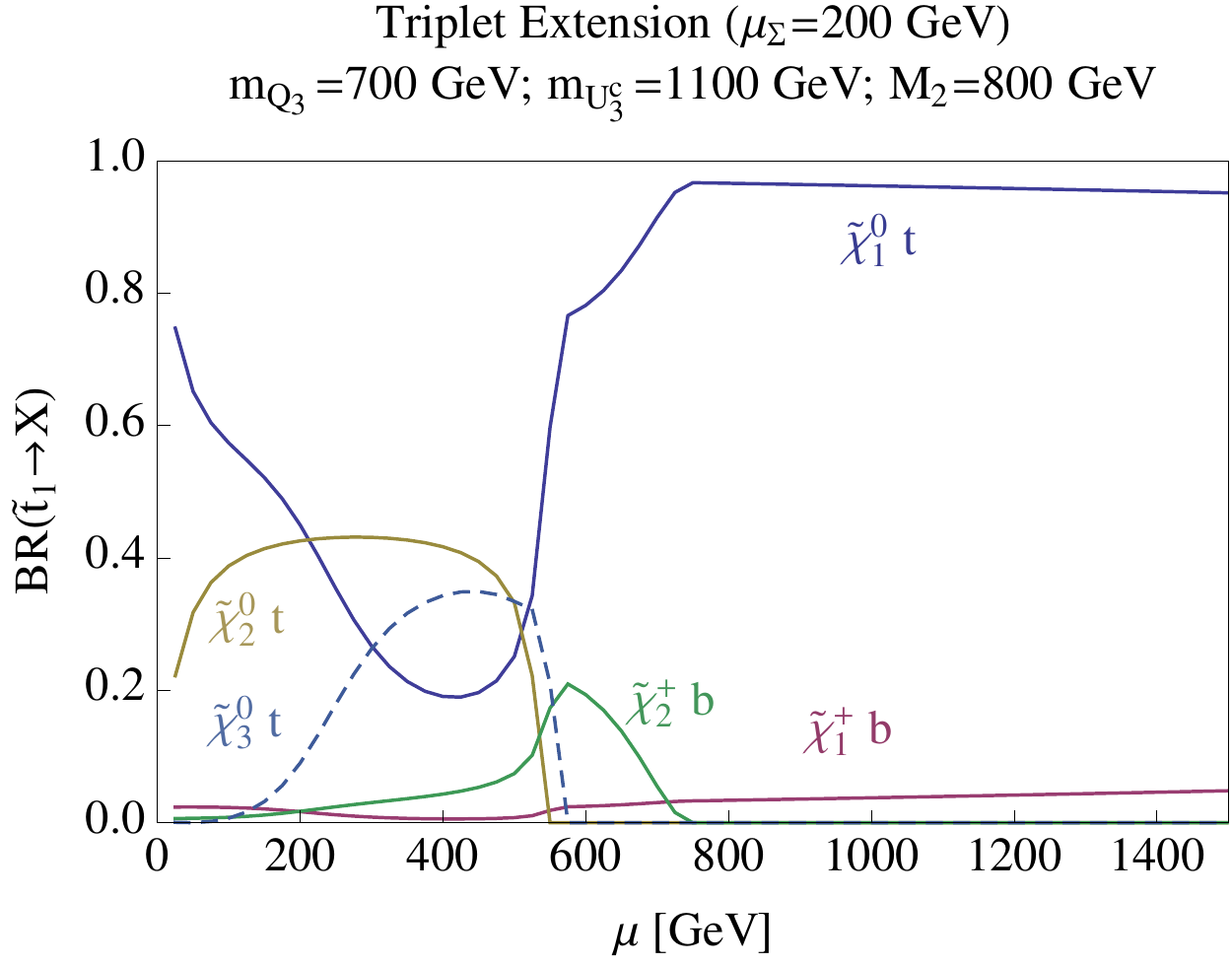}
                \caption{}
                \label{fig_pC_T}
        \end{subfigure}
        \quad
         \centering
         \begin{subfigure}[b]{0.48\textwidth}
                \centering
                \includegraphics[width=\textwidth]{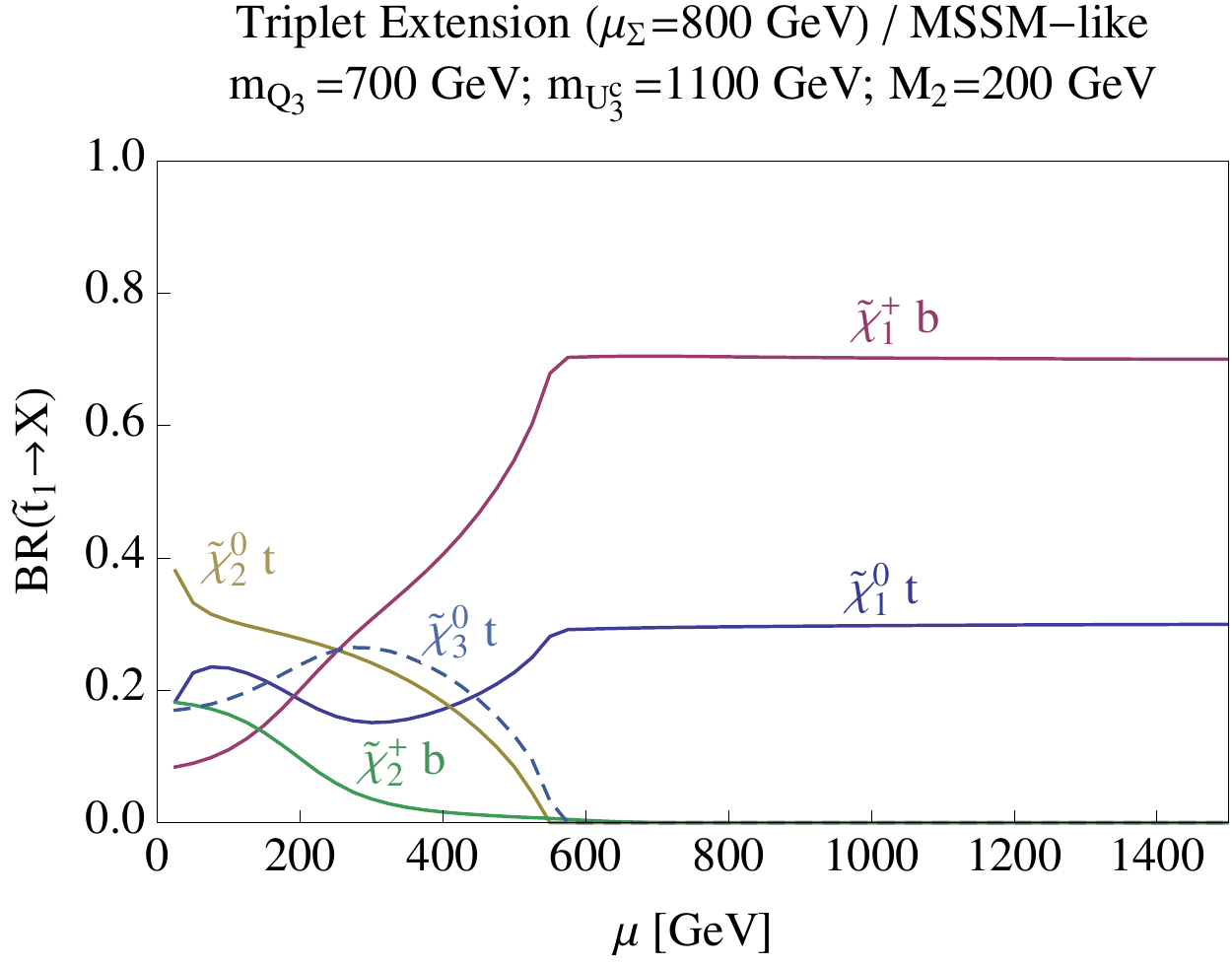}
                \caption{}
                \label{fig_pC_W}
        \end{subfigure}
        \caption{Comparison of the branching ratios of the lightest stop, as a function of $\mu$, for {\it point C}: $m_{Q_3}=700$ GeV, $m_{U_3^c}=1100$ GeV. (\subref{fig_pC_H}) In the triplet extension for $\mu_{\Sigma} = 800$ GeV (at the critical point) or, equivalently, the MSSM with $\tan\beta=3.48$. (\subref{fig_pC_T}) In the triplet extension with $\mu_{\Sigma} = 200$ GeV.  In this case the critical values are $(\tan\beta_C, \lambda_C)= (5.99,0.876)$.  (\subref{fig_pC_W}) The same in the MSSM-like case, with $\tan\beta=5.99$ and $M_2 = 200$ GeV so the stop also has a light triplet state available for decays. \label{fig_Point_C}}
\end{figure}

In other regions of the stop parameter space the presence of the fermionic triplets can have different effects, as we have verified scanning over different values of the stop soft masses $m_{Q_3}$ and $m_{U_3^c}$. Fig.~\ref{fig_Point_B} shows the analogous comparison for {\it point B}, which illustrates the case $m_{Q_3}\gg m_{U_3^c}$, giving a mostly right-handed lightest stop. Comparing Figs.~\ref{fig_pB_T} and \ref{fig_pB_W} (where $\lambda\simeq 0.97$ to reproduce the Higgs mass) we see that the effect of the triplet is the opposite as above, and the branching ratio to the second chargino is smaller compared to the MSSM-like scenario with a light Wino. The argument to understand this follows the one above. In this case, however, since the lightest stop is mostly right handed, and therefore has no Wino coupling, the suppression of the decays into $\tilde{\chi}^{0,\pm}_1$ is larger in the MSSM-like case. More interesting is the case $m_{U_3^c}\gg m_{Q_3}$, where the lightest stop is left-handed. The comparison of the different scenarios for {\it point C} are shown in Fig.~\ref{fig_Point_C}. In this case, Figs.~\ref{fig_pC_T} and \ref{fig_pC_W} (again $\lambda\simeq 0.97$ to reproduce a 125.5 GeV Higgs) show not only an enhancement in the decays to the second neutralino, but also a big effect on the branching fractions for the decays into the first neutralino/chargino states. This is more noticeable for large $\mu$, where only the decays into the lightest neutralino and chargino are available. As can be seen, the decays with the largest branching fraction are reversed, which completely changes the search strategy. This is solely explained by the lightest stop being left handed, so the only $\tilde{t}_1 \tilde{H}^\pm b$ interaction is proportional to the bottom Yukawa coupling. Thus, in the triplet extension charged decays are suppressed compared to the neutral ones by the ratio between the bottom and top Yukawa couplings. Again, in the MSSM-like scenario both occur through $SU(2)_L$ gauge couplings.

To summarize, the results in this section show that a light $Y=0$ fermion triplet state can significantly change the decays of the stops. As we will see in the next section, these changes translate into significant differences when we consider the full decay chains and analyze stop pair production in terms of states detectable at the detector level in LHC experiments.

\section{LHC considerations}
\label{Sec_LHC}

To perform a detailed search method for the range of parameters considered above is beyond the scope of this paper. However, we wish to highlight how the presence of a light fermion triplet could affect the final states observed at the detector level. To do this, we used FeynRules~\cite{Christensen:2008py, Duhr:2011se} to implement the model into MadGraph 5~\cite{Alwall:2011uj}. We generated stop pair production events at a center of mass energy of 14 TeV. The results were then ran through Pythia~\cite{Sjostrand:2006za} for hadronization and PGS~\cite{pgs} for detector level analysis.

We have analyzed all the scenarios discussed in the previous section: \textbf{(a)} a MSSM-like model with $M_2=800$ GeV (dubbed as \textit{MSSM$_{800}$});  \textbf{(b)} a triplet extension with $M_2=800$ GeV and $\mu_{\Sigma} = 200$ GeV (dubbed as \textit{Triplet Ext.}); and \textbf{(c)} another MSSM-like scenario with light Winos, $M_2=200\units{GeV}$ (dubbed as \textit{MSSM$_{200}$}). We chose a value of $\mu = 600$ GeV in all cases and generate events for all three points A, B and C in (\ref{points}). Within each scenario the masses of the first two neutralinos/charginos are very similar for all three points,
\begin{equation}
m_{\tilde{\chi}^0_{1,2}}\approx\left\{\begin{array}{c l}570,~\!600~\!\mathrm{GeV}&\mbox{MSSM}_{800}\\200,~\!580~\!\mathrm{GeV}&\mbox{Triplet Ext.}\\190,~\!600~\!\mathrm{GeV}&\mbox{MSSM}_{200}\end{array}\right.,~~~
m_{\tilde{\chi}^\pm_{1,2}}\approx\left\{\begin{array}{c l}580,~\!820~\!\mathrm{GeV}&\mbox{MSSM}_{800}\\200,~\!600~\!\mathrm{GeV}&\mbox{Triplet Ext.}\\190,~\!610~\!\mathrm{GeV}&\mbox{MSSM}_{200}\end{array}\right.,~~~
\end{equation}
while the stop masses and mixings are given by,
\begin{equation}
m_{\tilde{t}_{1,2}}\approx\left\{\begin{array}{c l}700,~\!740~\!\mathrm{GeV}&\mbox{Point A}\\720,~\!1110~\!\mathrm{GeV}&\mbox{Points B \& C}\end{array}\right.,~~~
\theta_{t}\approx\left\{\begin{array}{c l}46^\circ&\mbox{Point A}\\4^\circ&\mbox{Point B}\\1^\circ&\mbox{Point C}\end{array}\right..
\end{equation}
Again, we have set $A_t = 0$. Thus, the only source of stop mixing is the off-diagonal term $-v \mu y_t \cos\beta$ in the stop mass matrix. This term is enough to generate a large mixing in {\it point A}, due to the degeneracy of the soft masses in this case. For this point, since the two physical stops have similar masses we consider the production of both states. Moreover, because the masses are similar and the stop mixing angle $\theta_t\approx 45^\circ$, the decays of $\tilde{t}_{1}$ and $\tilde{t}_{2}$ are almost the same for this point.

In all cases the MSSM$_{800}$ scenario behaves as one of the simplified models under consideration at the LHC collaborations, where the lightest stop decays only to the lightest chargino and a bottom quark. In this situation the lightest chargino decays to the lightest neutralino with an off-shell $W$. On the other hand, from the results in the previous section, for the light fermion triplet scenario (Triplet Ext.~model)  we have:
\begin{equation}
\mbox{BR}({\tilde{t}_{1}\rightarrow\tilde{\chi}^0_{1}~\!t)}\approx\left\{\begin{array}{c l}0.2&\mbox{Points A \& B}\\0.8&\mbox{Point C}\end{array}\right.,~~~
\mbox{BR}({\tilde{t}_{1}\rightarrow\tilde{\chi}^\pm_{1,2}~\!b)}\approx\left\{\begin{array}{c l}0.3,~\!0.5&\mbox{Points A \& B}\\0\phantom{.0},~\!0.2&\mbox{Point C}\end{array}\right..
\label{BRTripletExt}
\end{equation}
There are several channels with sizable branching ratios or, in the case of {\it point C}, the decays are mostly dominated by a different channel  ($\tilde{t}_1\rightarrow \tilde{\chi}^0_1~\! t$) than in the MSSM  ($\tilde{t}_1\rightarrow \tilde{\chi}^+_1~\! b$). In any case, the phenomenology of the signal is expected to be different than in the MSSM.
When the stop decays to the second to lightest chargino, a long decay chain may be established. Note that for the splitting between the first and second chargino states, given the low value of the CP-odd Higgs mass, $m_A=130\units{GeV}$, the decays products of the second chargino also involve the heavy Higgses, except for the MSSM$_{200}$ where we are assuming a Higgs sector in the decoupling limit (i.e.~$m_A\simeq 1$ TeV). 

\begin{table}[t]
\caption{PGS detector level counts. The figure under every jet count is the percentage of the events which contain the given number of jets.}
\begin{center}
\begin{tabular*}{0.85\columnwidth}{@{\extracolsep{\fill}} c c c c c c}
\ctoprule
& &\multicolumn{4}{c}{$\#$ of Jet Counts} \\
Model &Point& 3 & 4 & 5 & $\ge6$ \\
\ctoprule
MSSM$_{800}$&A, B, C &	25 & 23-25	& 16-17	& 17-19  \\
$_{(M_2=800~\!\!\mathrm{GeV}})$& & & & &\\
\cmrule
Triplet Extension\TSspc&A, B, C& $\phantom{0}$8-9 & 14 & 18-19 & 54-57  \\
$_{(\mu_\Sigma=200~\!\!\mathrm{GeV},~M_2=800~\!\!\mathrm{GeV})}$\TSspc& & & & &\\
\cmrule
MSSM$_{200}$&A, B & 11-13 & 18-19 & 20-21 & 43-46  \\
$_{(M_2=200~\!\!\mathrm{GeV})}$& & & & &\\[-0.35cm]
&C & $\phantom{0}$7 & 13 & 17 & 60 \\
\cbottomrule
\end{tabular*}
\end{center}
\label{tab_Detector_Jets}
\end{table}%
\begin{table}[t]
\caption{PGS detector level counts. The figure under every lepton count is the percentage of the events which contain the given number of leptons.}
\begin{center}
\begin{tabular*}{0.85\columnwidth}{@{\extracolsep{\fill}} c c c c c c}
\ctoprule
& & \multicolumn{4}{c}{$\#$ of Lepton Counts} \\
Model  &Point&0 & 1 & 2 & $\ge3$ \\
\ctoprule
MSSM$_{800}$&A, B, C&	81-82& 17	& $\phantom{0}$1-2	&$\phantom{0}$0 \\
$_{(M_2=800~\!\!\mathrm{GeV}})$& & & & &\\
\cmrule
Triplet Extension\TSspc&A, B, C&  53-57& 34-37& $\phantom{0}$8-9& $\phantom{0}$1  \\
$_{(\mu_\Sigma=200~\!\!\mathrm{GeV},~M_2=800~\!\!\mathrm{GeV})}$\TSspc& & & & &\\
\cmrule
MSSM$_{200}$&A, B& 65& 30 & $\phantom{0}$5& $\phantom{0}$1 \\
$_{(M_2=200~\!\!\mathrm{GeV}})$& & & & &\\[-0.35cm]
&C& 53 & 35 & 10 & $\phantom{0}$2 \\
\cbottomrule
\end{tabular*}
\end{center}
\label{tab_Detector_Leptons}
\end{table}

The PGS detector level results for the jet and lepton counts are shown in Tables~\ref{tab_Detector_Jets} and~\ref{tab_Detector_Leptons}, respectively. These numbers include only new physics events $p p \rightarrow \tilde{t}_1^*\tilde{t}_1$ (and $\tilde{t}_2^*\tilde{t}_2$ for {\it point A}), without any cuts, and are only meant to illustrate the effects of the triplet. Looking at Eq.~(\ref{BRTripletExt}) we do not expect too much of a difference between the different benchmark points using this kind of simplistic analysis. However, we do see a big difference in the jet and lepton counts between the MSSM$_{800}$ compared to the Triplet Ext.~model and the MSSM$_{200}$.
 When there is a light triplet state ($\widetilde{\xi}^i$ or $\widetilde{W}^i$) we see that there is only a nominal difference between the Triplet Ext. and the MSSM$_{200}$  in the jet or leptons counts. One can imagine a more elaborate analysis to distinguish between both these scenarios. For example, in {\it point C}, most bottoms in the Triplet Ext.~model come from top decays, as opposed to the MSSM$_{200}$ where they come directly from the decay of the stop,
 as can be seen from Figs.~\ref{fig_pC_T} and \ref{fig_pC_W}. Therefore reconstructing the top mass could be a discriminant between both scenarios. 

\section{Conclusions}
\label{section_Conclusions}
The $Y=0$ triplet chiral field extension of the MSSM has many promising features.  The presence of extra particles allows the mass of the lightest Higgs boson to be raised to the observed 125.5 GeV value without the necessity of heavy stops or large mixing parameters. This helps to prevent the so called little hierarchy problem. In fact, the area of parameter space that we have worked in does not allow for a 125.5 GeV Higgs in the MSSM, but it is allowed with the addition of the triplet chiral superfield. This alone seems to be motivation enough for more study of this model. Moreover, the extra charged states coupling to the Higgs could explain a possible deviation in the $h\rightarrow\gamma\gamma$ channel. From the point of view of SUSY searches at the LHC, and in particular in stop pair production, the phenomenology can also be quite different, even if the triplet cannot couple directly to the stops. We have shown that fermion triplet states can introduce significant changes in the landscape of the stop decays via mixing in the neutralino/chargino sectors. These changes depend on the stop spectrum and in many cases can be as large as to modify the dominant decay modes of the lightest stop.
Should the LHC observe a supersymmetric like signal, a dedicated analysis looking for these kind of differences would allow to unveil the presence of triplet chiral superfields enlarging the MSSM spectrum.  

\section*{Acknowledgments}
This research was partly supported by the Notre Dame Center for Research Computing through computational resources. The work of MQ was supported in part by the European Commission under the ERC Advanced Grant BSMOXFORD 228169, by the Spanish Consolider-Ingenio 2010 Programme CPAN (CSD2007-00042) and by CICYT-FEDER-FPA2011-25948. We would also like to thank  B.~Fuks and  O.~Mattelaer  for assistance with FeynRules and MadGraph. JB, AD and BO were partly supported by the National Science Foundation under grant PHY-1215979. The work of JB has been also supported by the European Research Council under the European Union's Seventh Framework Programme (FP~\!\!/~\!\!2007-2013)~\!/~\!ERC Grant Agreement n. 279972.



\end{document}